\documentclass[pre,aps,twocolumn]{revtex4}

\usepackage{graphicx}
\usepackage{amsmath}
\usepackage{amssymb}
\usepackage{ifthen}
\usepackage{booktabs}

\usepackage{graphicx}
\usepackage{dcolumn}
\usepackage{bm}
\usepackage{longtable}
\usepackage{gensymb}
\usepackage{color}

\newcommand{\bb}{\begin{equation}}
\newcommand{\ee}{\end{equation}}
\newcommand{\ba}{\begin{eqnarray*}}
\newcommand{\ea}{\end{eqnarray*}}
\newcommand{\rhor}{\rho({\bf r})}
\newcommand{\dd}{{\rm d}}
\newcommand{\rr}{{\mathbf r}}
\newcommand{\dr}{{\rm d}{\bf r}}

\begin{document}

\title{Filling, depinning, unbinding:\\ Three adsorption regimes for nanocorrugated substrates}

\author{Alexandr \surname{Malijevsk\'y}}
\affiliation{
{Department of Physical Chemistry, University of Chemical Technology Prague, Praha 6, 166 28, Czech Republic;}\\
 {Department of Molecular and Mesoscopic Modelling, ICPF of the Czech Academy Sciences, Prague, 165 02, Czech Republic}}                

\begin{abstract}
\noindent We study adsorption at periodically corrugated substrates formed by scoring rectangular grooves into a planar solid wall which interacts
with the fluid via long-range (dispersion) forces. The grooves are assumed to be macroscopically long but their depth, width and separations can all
be molecularly small. We show that the entire adsorption process can be divided into three parts consisting of (i) \emph{filling} the grooves by a
capillary liquid; (ii) \emph{depinning} of the liquid-gas interface from the wall edges; and (iii) \emph{unbinding} of the interface from the top of
the wall, which is accompanied by a rapid but continuous flattening of its shape. Using a nonlocal density functional theory and mesoscopic
interfacial models all the regimes are discussed in some detail to reveal the complexity of the entire process and subtle aspects that affect its
behaviour. In particular, it is shown that the nature of the depinning phenomenon is governed by the width of the wall pillars (separating grooves),
whilst the grooves width only controls the location of the depinning first-order transition, if present.
\end{abstract}

\maketitle

\section{Introduction}

Wetting phenomena, i.e., intrusion of a liquid phase at a wall-gas interface, have been a subject of enormous scrutiny, both theoretical and
experimental, over the last few decades. In particular, for the simplest case of perfectly flat and chemically homogenous substrates there exist
several comprehensive reviews \cite{sullivan, dietrich, row, schick, forgacs, bonn} that summarize in detail the most fundamental aspects of wetting
transitions. Theoretical description of these phenomena relies on various approaches of different length scales. On a macroscopic level, the crucial
quantity is Young's contact angle $\theta$ of a macroscopic liquid droplet sitting on a substrate. A positive value of $\theta$ corresponds to
partial wetting states, while $\theta=0$ characterizes completely wet substrates and temperature $T_w$, at which the contact angle vanishes, is
called the wetting temperature. In a more microscopic approach, wetting states can be described in terms of a mean thickness $\ell$ of an adsorbed
liquid film, such that its value is microscopic below $T_w$ and macroscopic (effectively infinite) above $T_w$. This mesoscopic approach allows to
study the nature of the wetting transitions at $T_w$, which can be first-order or continuous, by inspecting the competition of the fluid-fluid and
wall-fluid microscopic forces. Furthermore, it also enables to describe other related phenomena such as complete wetting, which corresponds to a
divergence of $\ell$ along an isotherm $T>T_w$, as the chemical potential $\mu$ approaches its saturation value $\mu_{\rm sat}$ from below:
 \bb
 \ell\sim |\delta\mu|^{-\beta_{co}}\,,\;\;\; {\rm as}\;\;\delta\mu\to 0^-\,.
 \ee
Here $\delta\mu\equiv\mu-\mu_{\rm sat}$ and $\beta_{co}$ is the (non-universal) critical exponent characterizing the divergence of the liquid film
height; specifically, for systems where the interaction at long distances is dominated by non-retarded dispersion forces $\beta_{co}=1/3$.

More recently, an attention has been focused on structured substrates, in which case a number of additional interfacial phenomena occur \cite{nature,
ishino, quere, binder, saam, bruschi, borm}. For example, for sinusoidally shaped walls the complete wetting may be preceded by an unbending
transition, characterized by an abrupt flattening of the liquid-gas interface from the state at which the interface follows the shape of the wall,
and which occurs provided the wall amplitude exceeds a certain critical value \cite{santori, rascon, rejmer2000, rejmer2002, rejmer2007, rodriguez}.
This behavior is further enriched if complex fluids, such as nematic liquid crystals, are considered \cite{patricio1, patricio2, patricio3, rojas},
for which a sequence of re-entrant transitions has been observed. All these studies rely mainly on an analysis of interfacial Hamiltonian models that
prove to be extremely helpful in a description of wetting phenomena on structureless substrates. In a more microscopic manner, wetting properties of
nanoscopically  corrugated substrates have been investigated using  molecular based approaches, such as molecular dynamics or a classical density
functional theory (DFT) \cite{tas, jiang, zhang, singh, mal_depin, giac, li, egorov}. Recently,  complete wetting ($T>T_w$) of microscopically
corrugated substrates formed of a one-dimensional array of rectangular grooves, each of width $L_g$ and depth $D$, and separated by pillars of width
$L_p$, has been studied using DFT \cite{mal_depin}; it has been shown that in the particular case of $L\equiv L_g=L_p$, there exist two
\emph{molecularly} small values of the corrugation parameter $L$, $L_c^-$ and $L_c^+$, such that the system exhibits \emph{depinning} transition
within the interval of $L_c^-<L<L_c^+$ at which the liquid-gas interface detaches from the edges of the pillars and its height jumps by a finite
value. The depinning phase boundary terminates at the critical value of the corrugation parameter $L_c^+$,  such that the adsorption on the
corrugated substrate becomes similar to that of a planar wall for $L>L_c^+$. The other limit of the phase transition $L_c^-$ corresponds to the
minimal value of $L$ at which the depinning transition still exists and which occurs right at saturation, $\mu=\mu_{\rm sat}$. Therefore, below
$L_c^-$ the interface remains bound to the wall even at saturation which prevents complete wetting of the wall.

\begin{figure}
\includegraphics[width=8cm]{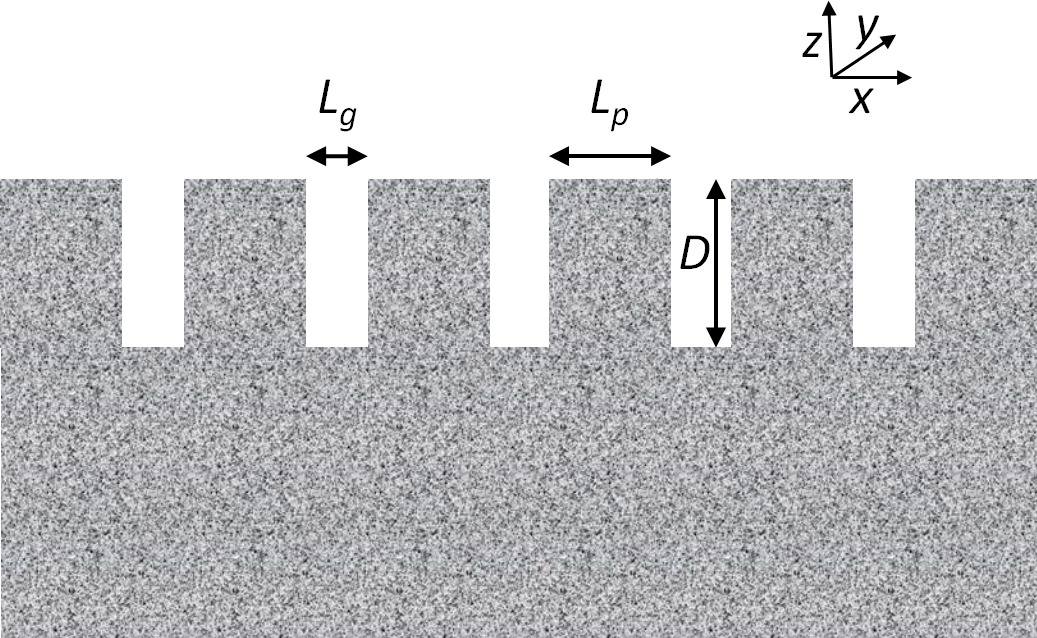}
\caption{Sketch of the substrate model in the $x$-$z$ projection of the Cartesian coordinate system. The wall consists of parallel grooves of width
$L_g$ and depth $D$ and are separated by pillars of width $L_p$. The grooves (and pillars) are assumed to be macroscopically long along the $y$-axis.
}
\end{figure}

The purpose of this work is to extend the previous study \cite{mal_depin} by generalizing the substrate model such that the parameters $L_g$ and
$L_p$ are now independent, which, in particular, allows to obtain a deeper understanding of the mechanism of the depinning transition. From a broader
perspective, however, we will show that the entire adsorption process exhibits a number of additional interfacial phenomena and  surface phase
transitions, and that it can be divided into three parts. The first regime corresponds to filling of the grooves with a dense, liquid-like phase, a
process which has recently attracted some attention \cite{bruschi,tas,darbellay,evans_cc,hofmann,
schoen,mal_groove,parry_groove,ser,mal_13,mistura13,our_groove,monson,fan,bruschi2, fin_groove_prl}. For a single macroscopically deep groove (or a
capped capillary), $D\to\infty$, the recent studies have shown that the filling is a first-order transition for temperatures below $T_w$ but
continuous (critical) otherwise with a critical exponent $\beta_g=1/4$ characterizing the rate of the groove filling for systems including dispersion
forces. In both cases, the transition occurs at the chemical potential $\mu_{cc}(L_g)$ corresponding to the location of the capillary condensation in
an infinite slit pore of width $L_g$.

However, for the current model the process of filling will be shown to somewhat deviate from these predictions in two aspects. Firstly, the grooves
depth $D$ considered here is not assumed to be generally macroscopic; for microscopic depths, the finite distance between the bottom and top of the
grooves brings about a competition between two effective repulsive forces pushing the liquid-gas interface away from both ends of the grooves in a
close analogy to condensation in slit pores formed of antisymmetric walls (contact angles $0$ and $\pi$) \cite{par_ev1, par_ev2, binder03, schulz,
stew, mal_as}. This analogy suggests that for temperatures higher than but not too far from $T_w$, there exists such a groove width $D^*(T)$ with
$T^*(D)\to T_w^+$ as $D\to\infty$,  for which the groove filling undergoes first-order localization-delocalization transition at $\mu_{cc}(L_g)$. In
this case, the low-adsorption state jumps to a state at which the grooves are filled with liquid to about a mid-height of the grooves.
Secondly, in contrast to the single groove model, the strength of the potentials exerted by the bottom and side walls inside each groove is now
different, since the side walls are no more semi-infinite. Consequently, the effective Hamaker constant of the side walls is now lower than that of
the bottom wall which means that for the current model  the filling process is expected to behave similarly to that in chemically heterogenous
grooves. Although this should not affect the order of the filling transition, the previous study on heterogenous grooves \cite{het_groove} suggests
that the singular behaviour of the critical filling transition for the current model will differ from that in an isolated groove.

As the second adsorption regime one can identify the process during which the liquid columns filling the grooves get connected. It will be shown that
the condition under which the process is continuous or discontinuous is determined solely by the pillar width $L_p$, while the groove width $L_g$
only affects the location $\mu_{\rm dep}$ of the depinning transition (if present), such that the dependence $\mu_{\rm dep}(L_g)$ is nonmonotonic.

The final adsorption regime corresponds to an unbinding of the liquid-gas interface from the top of the wall. The process is similar to complete
wetting of a planar wall and obeys the same power-law for the interface height in the limit of $\mu\to\mu_{\rm sat}$. However, the shape of the
interface which reflects the lateral inhomogeneity of the wall is now periodically undulated with amplitude $a$, which will be shown to decrease
continuously as the saturation is approached, such that $a\sim |\delta\mu|^{4/3}$, as $\delta\mu\to0$.

The rest of the paper is organized as follows. In section II we set the microscopic model by defining the fluid-fluid and wall-fluid interactions,
and formulate the DFT model based on Rosenfeld's fundamental measure theory. The numerical DFT results and analytic predictions of mesoscopic models
are presented in section III which is divided into three sub-sections, each devoted to one of the three adsorption regimes. Finally, summary of the
work, the discussion of the main results and outlook for extensions of the current study are subjects of section IV.

\section{Microscopic model}

Classical density functional theory \cite{evans79} is a statistical mechanical tool for a description of structure, thermodynamics, and phase
behaviour of inhomogeneous molecular fluids. The theory is based on a construction of a free energy functional $F[\rho]$ of  one-body fluid density
$\rhor$ which, except for some very particular cases, requires approximations. Specifically, for simple fluids one typically follows the perturbative
scheme in the spirit of the van der Waals theory:
 \bb
 F[\rho]=F_{\rm id}[\rho]+F_{\rm hs}[\rho]+F_{\rm att}[\rho]\,, \label{f_dft}
 \ee
 which splits the functional into the ideal gas, $F_{\rm id}$, repulsive hard-sphere, $F_{\rm hs}$, and attractive, $F_{\rm att}$, contributions.

 The ideal gas part is known exactly and is given by
  \bb
  \beta F_{\rm id}[\rho]=\int\dr\rho(\rr)\left[\ln(\rhor\Lambda^3)-1\right]
  \ee
  where $\Lambda$ is the thermal de Broglie wavelength and $\beta=1/k_BT$ is the inverse temperature.

The repulsive interaction of fluid molecules is mapped on the hard-sphere potential and its free energy contribution is approximated using
Rosenfeld's fundamental measure theory \cite{ros}
 \bb
F_{\rm hs}[\rho]=\frac{1}{\beta}\int\dd\rr\,\Phi(\{n_\alpha\})\,,\label{fmt}
 \ee
 where $\{n_\alpha\}$ denotes a set of six weighted densities
 \bb
 n_\alpha(\rr)=\int\dr'\rho(\rr')w_\alpha(\rr-\rr')\,;\;\;\alpha=\{0,1,2,3,v1,v2\}\,, \label{n_alpha}
 \ee
given by convolutions between one-body fluid density (or density profile) and the weight functions $w_\alpha$ which characterize so called
fundamental measures of the hard-sphere particles of diameter $\sigma$:
 \begin{eqnarray}
 w_3(\rr)&=&\Theta(R-|\rr|)\,,\;\;\;w_2(\rr)=\delta(R-|\rr|)\,,\\
 w_1(\rr)&=&w_2(\rr)/4\pi R\,,\;\;\,w_0(\rr)=w_2(\rr)/4\pi R^2\,,\\
 w_{v2}(\rr)&=&\frac{\rr}{R}\delta(R-|\rr|)\,,w_{v1}(\rr)=w_{v2}(\rr)/4\pi R\,.
 \end{eqnarray}
Here, $\Theta$ is the Heaviside function, and $\delta$ is Dirac's delta function and $R=\sigma/2$. Among various versions describing the free energy
density $\Phi$ for the inhomogeneous hard-sphere fluid, the original Rosenfeld prescription \cite{ros} was adopted, which accurately describes
short-range correlations between fluid particles and satisfies exact statistical mechanical sum rules \cite{hend}.

For separations $r>\sigma$, a pair of fluid particles are assumed to interact via the attractive part of the Lennard-Jones potential, which is
truncated at the cut-off of $r_c=2.5\,\sigma$:
 \bb
 u_{\rm a}(r)=\left\{\begin{array}{cc}
 0\,;&r<\sigma\,,\\
-4\varepsilon\left(\frac{\sigma}{r}\right)^6\,;& \sigma<r<r_c\,,\\
0\,;&r>r_c\,.
\end{array}\right.\label{ua}
 \ee
This attractive contribution is included to the free energy functional (\ref{f_dft}) in the usual mean-field fashion:
 \bb
F_{\rm att}[\rho]=\frac{1}{2}\int\int\dd\rr\dd\rr'\rhor\rho(\rr')u_{\rm a}(|\rr-\rr'|)\,.
 \ee

Having set the approximative free energy functional, the equilibrium density profile is obtained by minimizing the grand potential functional
 \bb
 \Omega[\rho]={\cal F}[\rho]+\int\dd\rr\rhor[V(\rr)-\mu]\,,\label{om}
 \ee
where $\mu$ is the chemical potential, and $V(\rr)$ is the external potential due to the static substrate (wall). For our model (see Fig.~1), the
latter can be written as
  \bb
 V(x,z)=V_\pi(z)+\sum_{n=-\infty}^\infty V_D(x+nP,z)\,, \label{wall_pot}
 \ee
where we have separated the contribution due to a planar wall $V_\pi(z)$ filling the volume $z<0$ and the potentials of the pillars, each of height
$D$ and width $L_p$ which are placed on the planar wall with a periodicity of $P=L_p+L_g$. We assume that the wall is formed by atoms that are
distributed uniformly with a density $\rho_w$ interacting with the fluid particles via the Lennard-Jones $12$-$6$ potential:
 \bb
 \phi_w(r)=4\varepsilon_w\left[\left(\frac{\sigma}{r}\right)^{12}-\left(\frac{\sigma}{r}\right)^{6}\right]\,. \label{phiw}
 \ee
Here we have identified the potential parameter $\sigma$ with the one for the fluid-fluid interaction (\ref{ua}) which we eventually use as a unit of
length, so that the strength of the potential is controlled by a single parameter $\varepsilon_w$. Integrating $\phi_w(r)$ over the half-space $z<0$
leads to the familiar Lennard-Jones $9$-$3$ potential of the planar wall:
 \bb
 V_\pi(z)=\left\{\begin{array}{cc}
 \pi\varepsilon_w\rho_w\sigma^3\left[\frac{1}{45}\left(\frac{\sigma}{z}\right)^9-\frac{1}{6}\left(\frac{\sigma}{z}\right)^3\right]\,;&z\ge0\,,\\
 \infty\,;&z<0\,.
 \end{array}\right. \label{vpi}
 \ee
The potential $V_D(x,z)$ is obtained by integrating $\phi_w(\rr)$ over the volume of a single pillar:
  \begin{eqnarray}
  V_D(x,z)&=&\rho_w\int_0^{L_p}\dd x'\int_{-\infty}^\infty\dd y'\int_0^{D}\dd z'\\
  &&\times\phi_w\left(\sqrt{(x-x')^2+y'^2+(z-z')^2}\right)\,,\nonumber
  \end{eqnarray}
which is valid everywhere except for the region inside the pillar where the potential is infinite. The integration can be carried out separately for
the attractive and the repulsive bits of $\phi_w$ which allows to split the potential as follows:
 \bb
 V_D(x,z)=V_6(x,z)+V_{12}(x,z)
 \ee
 where
 \begin{eqnarray}
V_6(x,z)&=&-\frac{\pi}{3}\varepsilon_w\sigma^6\rho_w\left[\psi_6(x,z)-\psi_6(x,z-D)\right.\nonumber\\
&&\left.-\psi_6(x-L_p,z)+\psi_6(x-L_p,z-D)\right] \label{v6}
 \end{eqnarray}
 and
 \begin{eqnarray}
V_{12}(x,z)&=&\pi\varepsilon_w\sigma^{12}\rho_w\left[\psi_{12}(x,z)-\psi_{12}(x,z-D)\right.\\
&&\left.-\psi_{12}(x-L_p,z)+\psi_{12}(x-L_p,z-D)\right]\,.\nonumber
 \end{eqnarray}
 Here,
  \bb
\psi_6(x,z)=\frac{2x^4+x^2z^2+2z^4}{2x^3z^3\sqrt{x^2+z^2}} \label{psi6}
 \ee
 and
  \begin{widetext}
 \bb
\psi_{12}(x,z)=\frac{1}{128}{\frac {128\,{x}^{16}+448\,{x}^{14}{z}^{2}+560\,{x}^{12}{z}^{4}+280\,{
x}^{10}{z}^{6}+35\,{x}^{8}{z}^{8}+280\,{x}^{6}{z}^{10}+560\,{x}^{4}{z} ^{12}+448\,{z}^{14}{x}^{2}+128\,{z}^{16}}{{z}^{9}{x}^{9} \left( {x}^{2
}+{z}^{2} \right) ^{7/2}}} -\frac{1}{z^9}\,.
 \ee
 \end{widetext}

Minimization of (\ref{om}) leads to the Euler-Lagrange equation
 \bb
 V(\rr)+\frac{\delta{\cal F}_{\rm hs}[\rho]}{\delta\rho(\rr)}+\int\dd\rr'\rho(\rr')u_{\rm a}(|\rr-\rr'|)=\mu\,,\label{el}
 \ee
which is solved iteratively on equidistantly discretized two dimensional grid with the spacing of $0.1\,\sigma$. The bulk properties of the fluid
model defined by Eq.~(\ref{ua}) is obtained by solving Eq.~(\ref{el}) by setting $V(\rr)=0$. This allows to construct the phase diagram of the bulk
fluid which terminates at the critical point at the temperature corresponding to $k_BT_c/\varepsilon=1.41$.

\section{Results}

\begin{figure}
\centerline{\includegraphics[width=8cm]{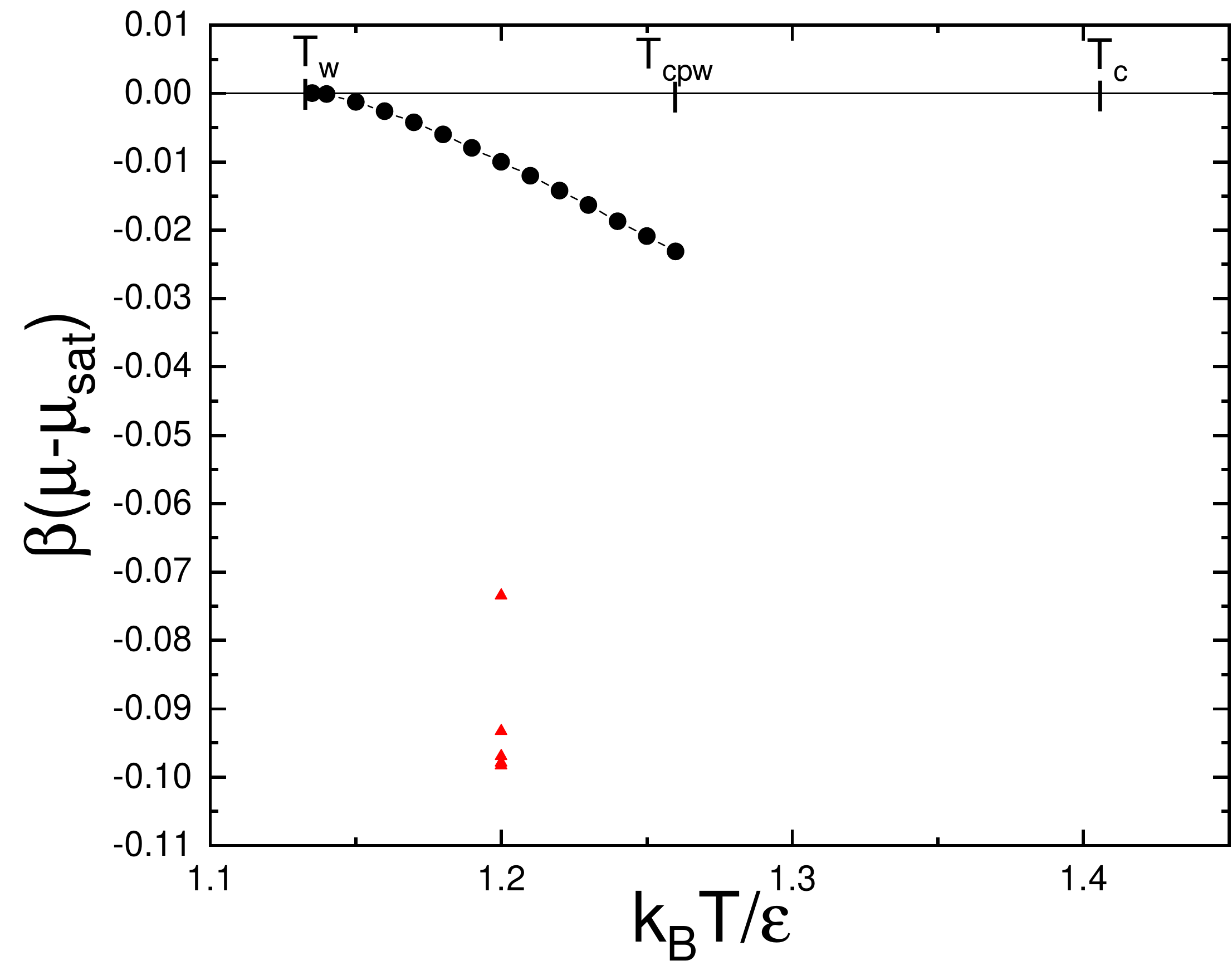}}
 \caption{Surface phase diagram for a planar wall interacting with the fluid via the potential given by Eq.~(\ref{vpi})
 with the strength parameter $\varepsilon_w=\varepsilon$. The bulk liquid-vapour coexistence corresponds to the horizontal line $\delta\mu=0$, on
 which the wetting temperature $T_w$, the critical prewetting temperature $T_{cpw}$ and the bulk critical temperature $T_c$ are denoted. The phase boundary
 for the prewetting transition connects the saturation line tangentially at $T_w$ and terminates at the critical prewetting point. Also shown are the
 values of the chemical potential departure from saturation, $\delta\mu=\mu-\mu_{\rm sat}(T)$, which correspond to capillary condensation
 (at temperature $k_BT/\varepsilon=1.2$) in infinite slits formed of a pair of walls a distance $L_g=10\,\sigma$ apart and of width (from top to bottom):
 $L_p=2\,\sigma$, $L_p=5\,\sigma$ $L_p=10\,\sigma$ $L_p=20\,\sigma$, and $L_p=\infty$.}\label{spd}
\end{figure}

Throughout this work, the strength of the wall potential defined in the previous section is fixed to $\varepsilon_w=\varepsilon$; the surface phase
diagram of the corresponding planar wall is displayed in Fig.~\ref{spd}. Since in our model the fluid-fluid interaction (\ref{ua}) is truncated and
thus effectively short-range as opposed to the long-range wall-fluid interaction, the wetting transition occurring at the temperature $T_w=0.8\,T_c$
and the bulk liquid-vapour coexistence is inevitably of first-order \cite{schick}. Consequently, also present is the prewetting line denoting the
phase boundary between thin and thick wetting layers which extends the surface free energy singularity at $T_w$ off bulk coexistence and terminates
at the critical prewetting temperature $T_{cpw}=0.88\,T_c$. Hereafter, in a study of complete wetting of periodically structured walls, we will
consider two isothermal paths with $T>T_w$, such that one,  $T=0.85\,T_c$ ($k_BT/\varepsilon=1.2$), crosses the prewetting line while the other,
$T=0.92\,T_c$ ($k_BT/\varepsilon=1.3$), does not.

\subsection{Filling}

 \begin{figure*}
\centerline{\includegraphics[width=8cm]{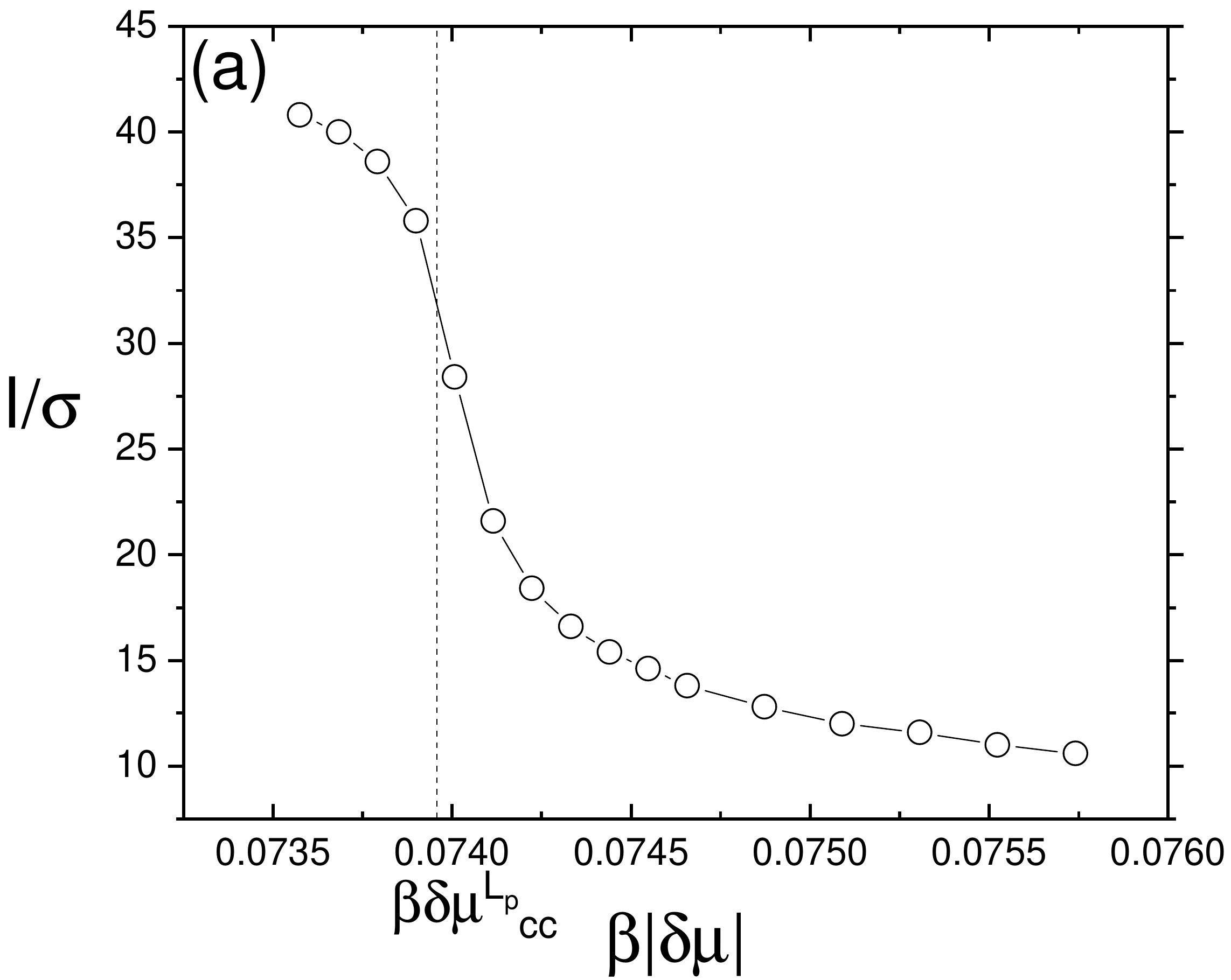}\hspace{0.5cm}\includegraphics[width=8cm]{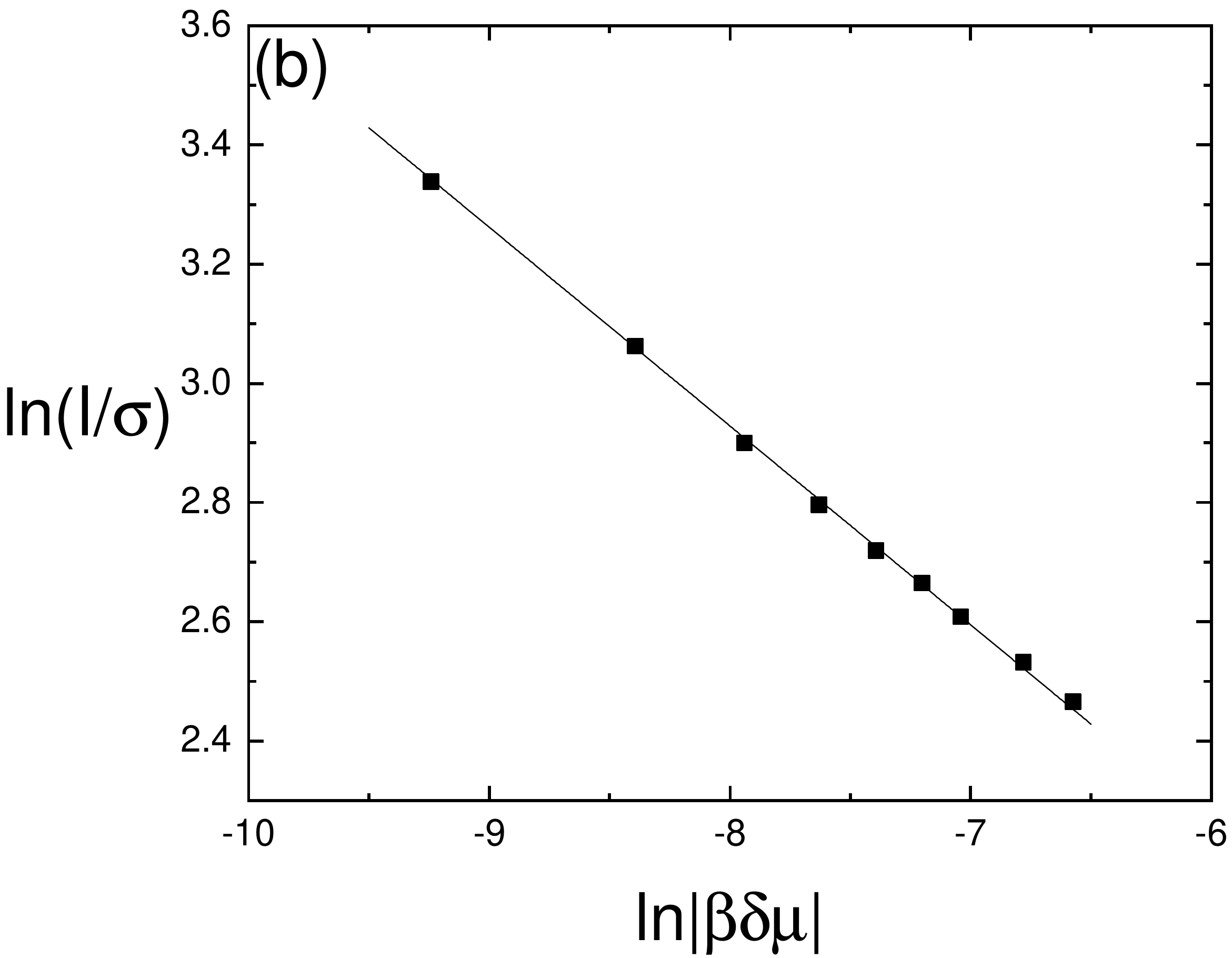}}
 \caption{Left: Meniscus height as a function of
the chemical potential departure from saturation $\delta\mu=\mu-\mu_{\rm sat}$ in a capillary groove of width $L_g=10\,\sigma$ and depth
$D=50\,\sigma$. The width of the side walls is $L_p=2\,\sigma$. The vertical dotted line denotes the chemical potential $\mu_{cc}^{L_p}(L_g)$
corresponding to the capillary condensation in an infinite pore formed by parallel walls of thicknesses $L_p$. Right: The log-log plot of the
previous dependence for $\mu<\mu_{cc}$. The line has a slope $-1/3$. For temperature $T=0.85\,T_c$.}\label{beta_g}
\end{figure*}

We start by discussing the first adsorption regime, i.e. the filling of grooves by a high density liquid-like phase. For $T>T_w$, a meniscus
separating liquid and gas phases in each groove is formed and the filling process can be described by monitoring the growth of the meniscus height as
the chemical potential is increased to the value near $\mu_{cc}(L_g)$. Here, a focus will be made on aspects specific for the current substrate model
which change some features of the filling process when compared to that occurring in a deep isolated groove, as already studied in detail. To this
end, the implications of finiteness of the side walls (pillars) as well as the depth of the grooves will be discussed. In this subsection, we will
consider the temperature $T=0.85\,T_c<T_{cpw}$ which allows for a prewetting jump and the filling inside of deep and shallow grooves will be
discussed separately.

\begin{figure}
\centerline{\includegraphics[width=5cm]{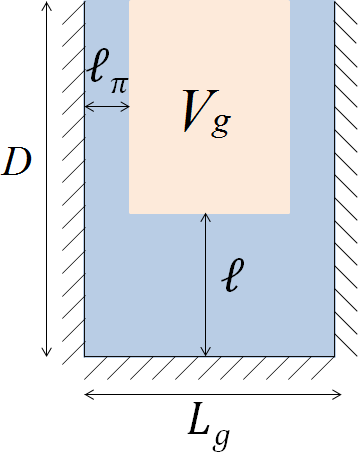}}
 \caption{Illustration of the sharp-kink approximation applied for the analysis of filling in a single groove.}\label{ska}
\end{figure}

\subsubsection{Deep grooves}

Fig.~\ref{beta_g}a displays DFT results for the meniscus height dependence on the chemical potential difference from saturation,
$\delta\mu=\mu-\mu_{\rm sat}$, for the substrate with deep grooves ($D=50\,\sigma$); the grooves are of  width of $L_g=10\,\sigma$ and are separated
by thin pillars of  width of $L_p=2\,\sigma$. For sufficiently low values of $\mu$ the meniscus height is largely determined by an effective
repulsion from the groove bottom which gives rise to a steep growth of the meniscus  as a certain threshold value of the chemical potential is
approached. It is well known that for single and macroscopically deep grooves the threshold value corresponds to $\mu_{cc}(L_g)$ pertinent to
capillary condensation in the infinite slit of the same width formed by two semi-infinite solid slabs. However, in this substrate model the grooves
are formed of side walls of finite thickness and the threshold now corresponds to the chemical potential $\mu_{cc}^{L_p}(L_g)$, rather than to
$\mu_{cc}(L_g)$, locating the capillary condensation in slits formed by solid slabs of finite thickness $L_p$. In the surface phase diagram of
Fig.~\ref{spd} the values $\delta\mu_{cc}^{L_p}(L_g)=\mu_{cc}^{L_p}(L_g)-\mu_{\rm sat}$ determining the location of the capillary condensation for
slits with $L_g=10\,\sigma$ and various widths of the confining walls obtained from DFT are depicted. We can see that  $\mu^{L_p}_{cc}(L_g)$ converge
towards $\mu_{cc}(L_g)$ rather rapidly and become hardly distinguishable from the limiting value already for $L_p\approx10\,\sigma$. However, for the
thin walls with $L_p=2\,\sigma$ the difference is pronounced and the value $\delta\mu_{cc}^{L_p}(L_g)$ does indeed correspond to the threshold shown
in Fig.~\ref{beta_g}a beyond which the growth of the meniscus height is considerably slower due to a strong effective repulsion acting from the
groove top \cite{our_groove}.

According to the previous DFT studies of single grooves \cite{mal_groove, our_groove}, the considered groove depth ($D=50\,\sigma$) is sufficient to
determine the critical exponent $\beta_g=1/4$ associated with the divergence of the interface height as $\delta\mu\to0^-$, as expected for infinitely
deep (and $L_p$ semi-infinite) grooves. However, the log-log plot of the dependence shown in Fig.~\ref{beta_g}b reveals that although the interface
height still satisfies the power law $\ell\sim(\mu_{cc}^{L_p}(L_g)-\mu)^{-\beta_g}$, the critical exponent is now $\beta_g=1/3$. The change in the
value of $\beta_g$ can be explained using the sharp-kink analysis based on the model sketched in Fig.~\ref{ska}. Here, we consider a single groove of
depth $D$ and width $L_g$ filled with a liquid of a constant density $\rho_l$ up to the height $\ell$ from the groove bottom. Moreover, we also
consider wetting layers of the width of $\ell_\pi$ adsorbed at both side walls of the groove (above the level of $\ell$) which should be taken into
account since the temperature of the system is $T>T_w$. The rest of the groove with a volume of $V_g=(L_g-2\ell_\pi)(D-\ell)$ is occupied by a gas of
density $\rho_g$. Within this approximation the excess grand potential functional per unit length relative to the system filled entirely by gas
reduces to a function of a single parameter $\ell$:
 \begin{eqnarray}
 \Omega(\ell)&=&(p-p_l^+)[L_g\ell+2\ell_\pi(D-\ell)]-2(\ell+\ell_\pi)\gamma\nonumber\\
 &&+(\rho_g-\rho_l)\int_{V_g} V_{\rm att}(\rr)\,,
 \end{eqnarray}
where $p$ is the pressure of the gas reservoir, $p_l^+$ is the pressure of the corresponding (same temperature and chemical potential) metastable
liquid and the last term expresses the effective interaction (binding potential) between the wall-liquid and liquid-gas interfaces where only the
attractive forces were included:
 \bb
 V_{\rm att}=\frac{2\alpha}{z^3}+V_6(x,z)+V_6(L_g-x,z)\,,
 \ee
 with $\alpha=-\pi\varepsilon_w\rho_w\sigma^3/3$ and $V_6(x,z)$ given by Eq.~(\ref{v6}).

The equilibrium state corresponds to the minimum of $\Omega(\ell)$ implying
 \begin{eqnarray}
 &&\delta\mu\Delta\rho(L-2\ell_\pi)+2\gamma=\Delta\rho\int_{\ell_\pi}^{L_g-\ell_\pi} V_{\rm att}(x,\ell)\,dx=\nonumber\\
 &=&\frac{2\alpha\Delta\rho(L_g-2\ell_\pi)}{\ell^3}+2\Delta\rho\int_{\ell_\pi}^{L_g-\ell_\pi} V_6(x,\ell)\,dx \label{dmu}
 \end{eqnarray}
 after $(p-p_l^+)\approx-\delta\mu\Delta\rho$ and $\Delta\rho=\rho_l-\rho_g$ have been substituted and the contribution due the bottom wall
 carried out.

Now, in the limit of $D\to\infty$ \emph{and} $L_p\to\infty$, the contribution to the binding potential from the side walls is
$2\alpha\Delta\rho(1/\ell_\pi^2-(L_g-2\ell_\pi)/\ell^3+\cdots)$. Furthermore, assuming that the film thickness $\ell_\pi$ at side walls is the same
as that for a planar wall, we can substitute $\ell_\pi=(2\alpha/\delta\mu)^{1/3}$ which leads to
 \begin{eqnarray}
 \delta\mu\Delta\rho(L_g-3\ell_\pi)+2\gamma&=&2\alpha\Delta\rho\frac{L_g-2\ell_\pi}{\ell^3}\label{dmu2}\\
 &&-2\alpha\Delta\rho
 \left[\frac{L_g-2\ell_\pi}{\ell^3}+{\cal{O}}(\ell^{-4})\right]\nonumber\,.
 \end{eqnarray}
Dividing by $\Delta\rho(L_g-3\ell_\pi)$ and using Kelvin's equation with Derjaguin's correction for  $\mu_{cc}(L_g)$ at $T>T_w$ \cite{derj, evans90}:
 \bb
 \mu_{cc}(L_g)=\mu_{\rm sat}-\frac{2\gamma}{\Delta\rho(L_g-3\ell_\pi)}\,,
 \ee
 Eq.~(\ref{dmu2}) implies
 \bb
 \delta\mu_{cc}(L_g)\equiv\mu-\mu_{cc}(L_g)\sim\ell^{-4}\;\;\;(L_p\to\infty)\,,
 \ee
and therefore $\beta_g=1/4$. However, it is straightforward to show that for $L_p$ finite, the side walls contribute to the binding potential with
$2\alpha\Delta\rho(1/\ell_\pi^2+{\cal{O}}(\ell^{-4}))$ which also follows from dimensional arguments, since the integration domain of the wall
potential along the $x$-axis is now finite. Therefore, the bottom wall contribution of the order of $\ell^{-3}$ is no more compensated and thus
 \bb
 \delta\mu_{cc}(L_g)\sim\ell^{-3}\;\;\;(L_p\;\;\mbox{finite})\,, \label{dmtret}
 \ee
 hence $\beta_g=1/3$.

Finally, for our model consisting of a periodic array of grooves, the contribution to the binding potential from the pillars will be of the same form
as for an isolated groove with $L_p$ infinite, i.e. $2\tilde{\alpha}\Delta\rho(1/\ell_\pi^2-(L_g-2\ell_\pi)/\ell^3+\cdots)$, but with a different
coefficient $\tilde{\alpha}$, since the effective potential strength of the pillars is now weaker. Therefore, the leading order term
${\cal{O}}(\ell^{-3})$ induced by the bottom wall does not cancel, which leads to Eq.~(\ref{dmtret}) again, and thus the critical coefficient remains
$\beta_g=1/3$, as for a single groove with side walls of finite width.

Although the value of the critical exponent $\beta_g$ is in line with our DFT results, it should be noted that according to the sharp-kink analysis
the steep growth in the height of the meniscus takes place in the vicinity of $\mu_{cc}(L_g)$ as given by the modified Kelvin equation, rather than
at $\mu_{cc}^{L_p}(L_g)$ as obtained from DFT. This microscopic-scale difference is clearly beyond the scope of the mesoscopic analysis and, in
particular, can be attributed to the fact that the approximation $\ell_\pi=(2\alpha/\delta\mu)^{1/3}$ becomes inaccurate for thin walls.

\subsubsection{Shallow grooves}

\begin{figure}
\centerline{\includegraphics[width=8cm]{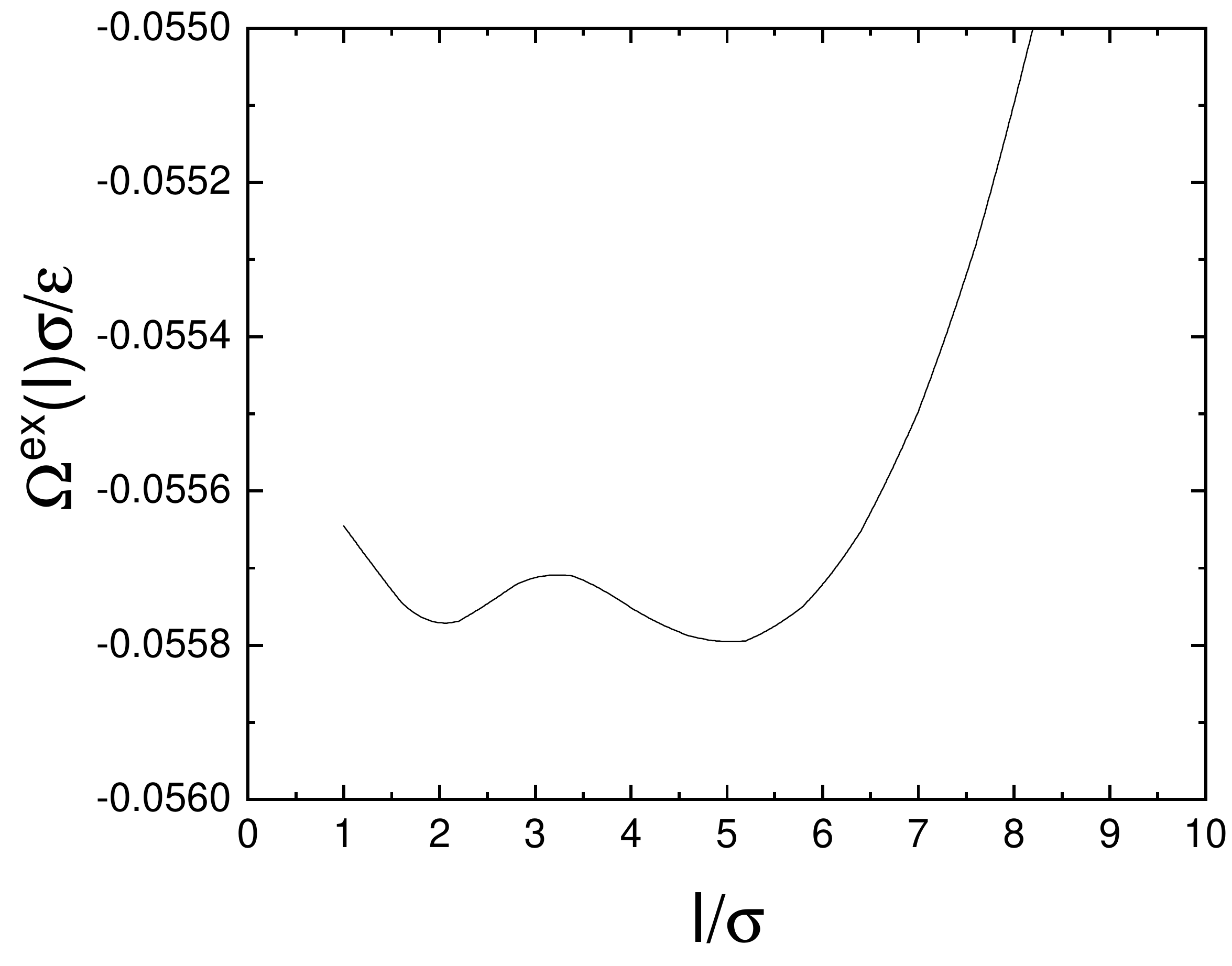}}
 \caption{Binding potential obtained from DFT by constrained minimization. For the substrate with the parameters $D=10\,\sigma$,
 $L_g=10\,\sigma$ and $L_p=2\,\sigma$, temperature $T=0.85\,T_c$ and the chemical potential $\beta\delta\mu=-0.093$.  }\label{bind_pot}
\end{figure}

\begin{figure}
\centerline{\includegraphics[width=4.5cm]{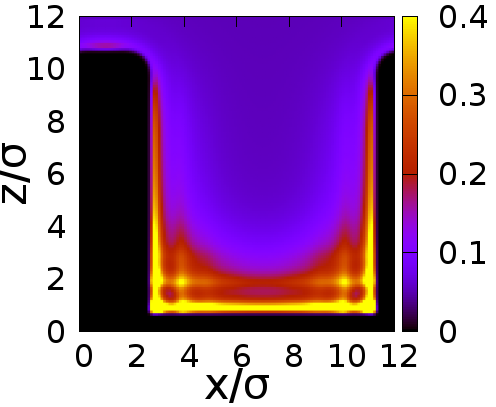} \includegraphics[width=4.5cm]{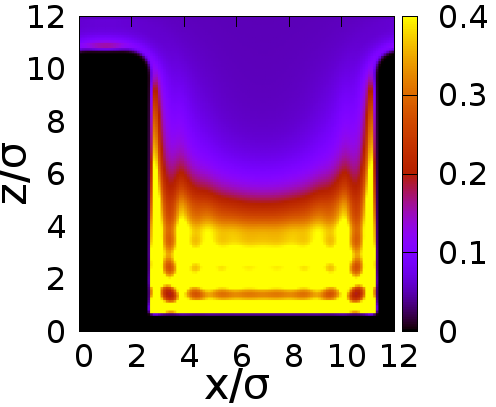}}
 \caption{Two-dimensional density profiles of two coexisting equilibrium states corresponding to the minima of the binding potential
 shown in Fig.~\ref{bind_pot}. Left panel: the low adsorption or localized state
 where the meniscus is pinned at the groove bottom. Right panel: delocalized state where the meniscus
 is near the mid-height of the groove. }\label{loc}
\end{figure}

Let us now consider the case when the depth $D$ of the grooves is microscopically small. The meniscus is now a subject of simultaneous effective
interactions acting from the groove bottom and the groove top which repel the meniscus from both ends of the groove. Such a system is thus
reminiscent of a slit pore formed of competing walls, as originally pointed out in Ref.~\cite{evans_cc} and explicit mapping between the two systems
has been made in Ref.~\cite{mal18}. Following this analogy one expects that a localization-delocalization first-order transition occurs along the
\emph{capillary} liquid-gas equilibrium line, i.e. at $\mu_{cc}(L_g;T)$ (or $\mu_{cc}^{L_p}(L_g;T)$ for finite walls), for the groove depth $D^*(T)$
which increases with decreasing $T$ with $D^*(T_w)\to\infty$, in which case the transition coincides with the groove filling \cite{mal_as}.  At the
same time, there should be a minimal value of the groove depth $D_c=D^*(T_s)$ allowing for the transition, where $T_s>T_w$ is the spinodal
temperature characterized by a disappearance of the energy barrier in the binding potential.

In order to test these arguments we present the DFT results for the groove with the same parameters as in the previous part, except that we now vary
the depth of the groove. At the fixed temperature of $T=0.85\,T_c$ and the capillary liquid-gas coexistence $\mu_{cc}^{L_p}(L_g;T)$, we construct the
constrained grand potential per unit length by fixing a certain value $\ell$ of the meniscus height from the groove bottom. We found that the
constrained grand potential possesses two equally deep minima for the groove depth of $D=10\,\sigma$ as is displayed in Fig.~\ref{bind_pot}. The
density profiles of the two coexisting states are shown in Fig.~\ref{loc}.

\subsection{Depinning}

\begin{figure}
\centerline{\includegraphics[width=8cm]{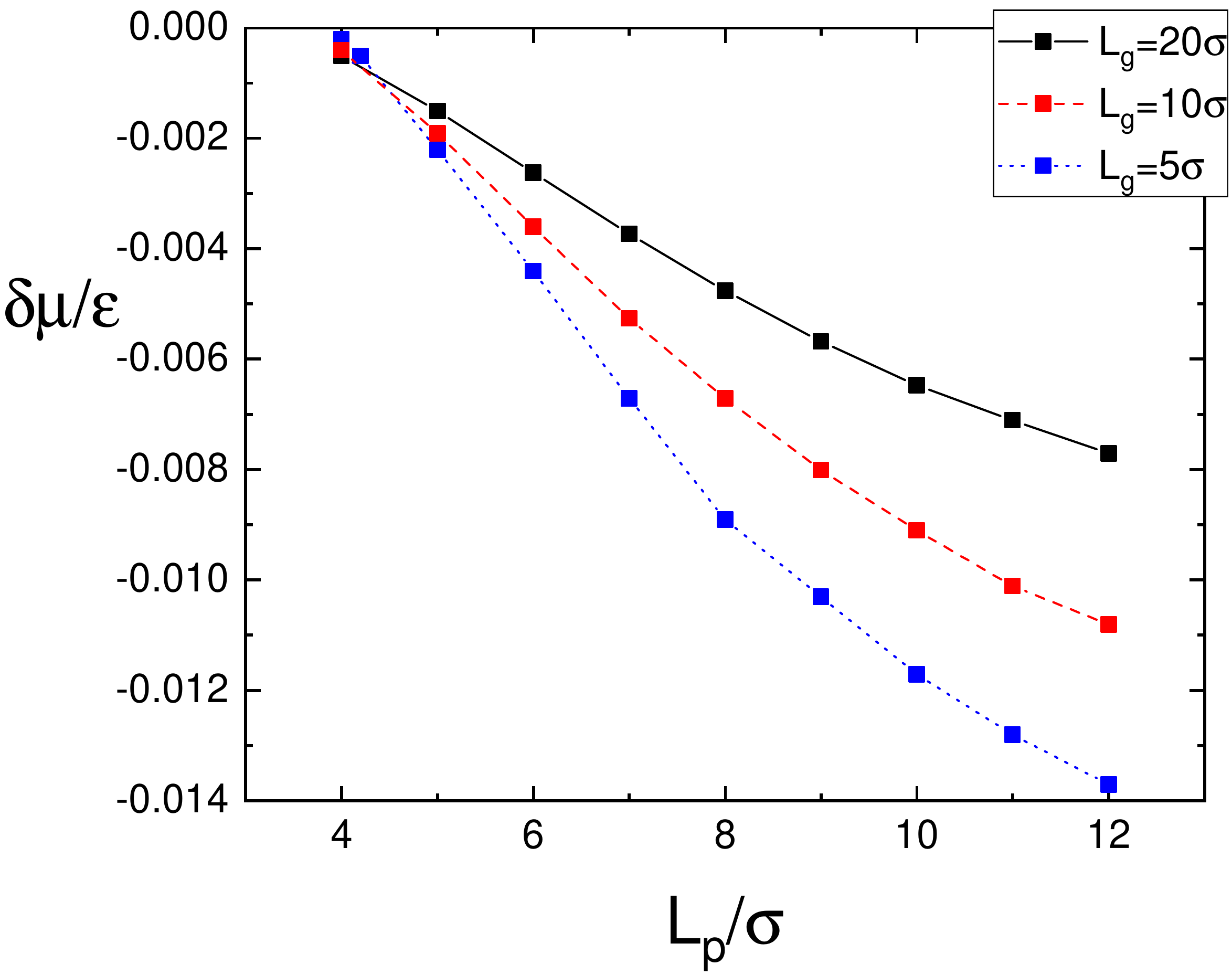}}
 \caption{The phase diagram of the depinning transition for three fixed values of the groove width. The critical point of the transition is about
 $L_{cp}^+\approx12\,\sigma$ and the depinning line connects the saturation line at $L_{cp}^-\approx4\,\sigma$ independently of the
 groove width. For $T=0.92\,T_c$. }\label{pd_L1}
\end{figure}

We now turn out attention to the second adsorption regime which we associate with merging of the liquid columns adsorbed in the grooves. As shown
previously for the special case of $L=L_g=L_p$, there exists a depinning transition for a sufficiently small value of $L$ \cite{mal_depin}. Within
the present model we wish to obtain a deeper understanding of the nature of the transition by investigating the roles of the parameters $L_g$ and
$L_p$ independently. To this end, we show in Fig.~\ref{pd_L1} the phase diagram of the depinning transition in the $L_p$--$\delta\mu$ plane for three
fixed values of the grooves width, $L_g=5\,\sigma$, $L_g=10\,\sigma$, and $L_g=20\,\sigma$, as obtained from DFT. The main upshot based on these
results is that the behaviour of the depinning transition is largely controlled by the pillars width $L_p$ with the critical values $L_{cp}^+$ and
$L_{cp}^-$, whose meanings are analogous to $L_c^+$ and $L_c^-$, and which seem to be independent on the grooves width. The only impact of $L_g$ is
that the phase boundary is shifted closer towards the coexistence line as $L_g$ increases.

\begin{figure}
\centerline{\includegraphics[width=8cm]{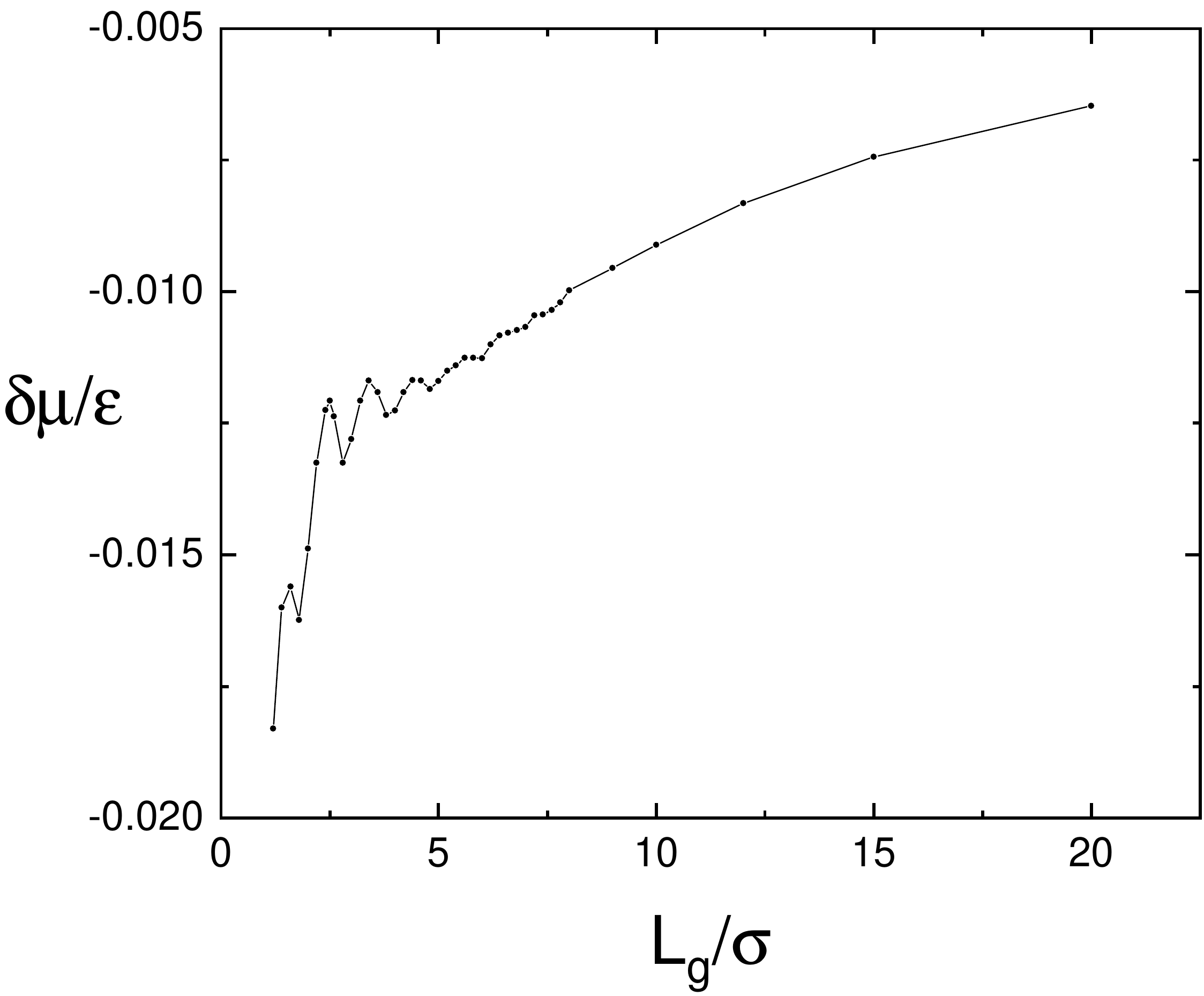}}
 \caption{The phase diagram of the depinning transition for the fixed value $L_p=10\,\sigma$ of the pillar width. The transition line is shown up to
 $L_g=20\,\sigma$ but proceeds onwards while it terminates at $L_g=1\,\sigma$ when no liquid can intrude the grooves.
 For $T=0.92\,T_c$.}\label{pd_L2}
\end{figure}

This is to a greater detail illustrated in Fig.~\ref{pd_L2} where we fix the pillar width, $L_p=10\,\sigma$, and vary the groove width $L_g$ instead.
Here, in contrast to the previous case, the line of the depinning transition is unbounded from above approaching asymptotically the bulk coexistence,
similarly to the case of capillary condensation, which, wherever present, of course precedes the depinning transition. In fact, over the displayed
range of $L_g$ the modulus of $\delta\mu$ corresponding to the capillary condensation (for the slit width of $L_g$ and the same temperature) is by
one order in magnitude larger than that for the depinning. Also, the capillary critical point at this temperature is about $L_c\approx6\,\sigma$,
while the depinning line terminates only at $L_g\approx\sigma$ in which case the fluid atoms cannot intrude the grooves anymore due to excluded
volume effects. Note that for even smaller values of $L_g$ the depinning transition would be replaced by the bridging transition on a planar but
chemically heterogenous wall consisting of periodically repeating hydrophilic and hydrophobic stripes \cite{posp19}.

\begin{figure}[h]
\includegraphics[width=4.0cm]{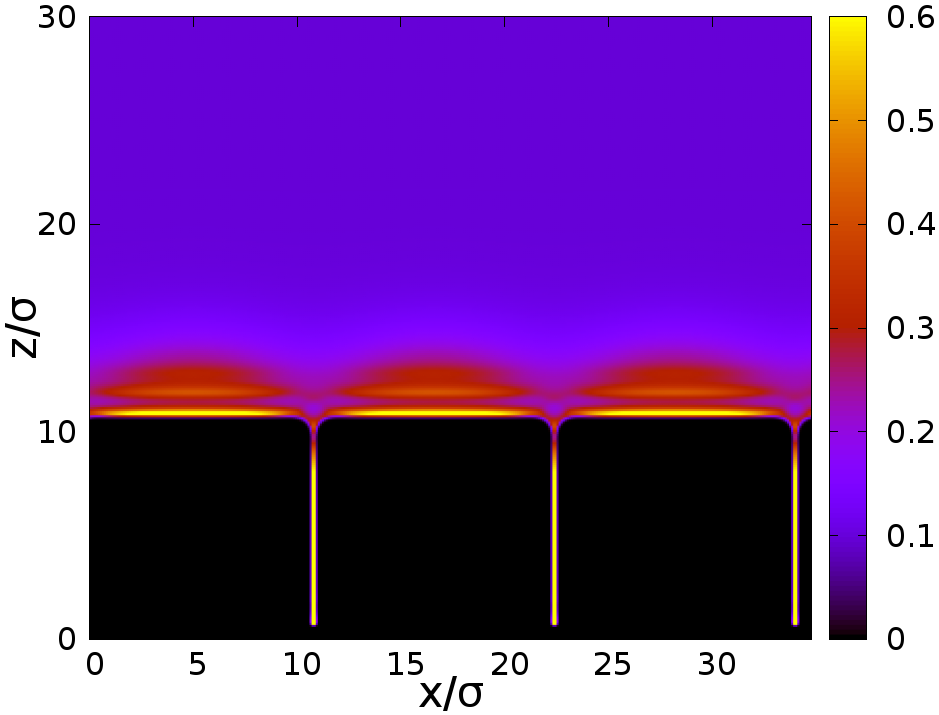} \hspace*{0.1cm} \includegraphics[width=4.0cm]{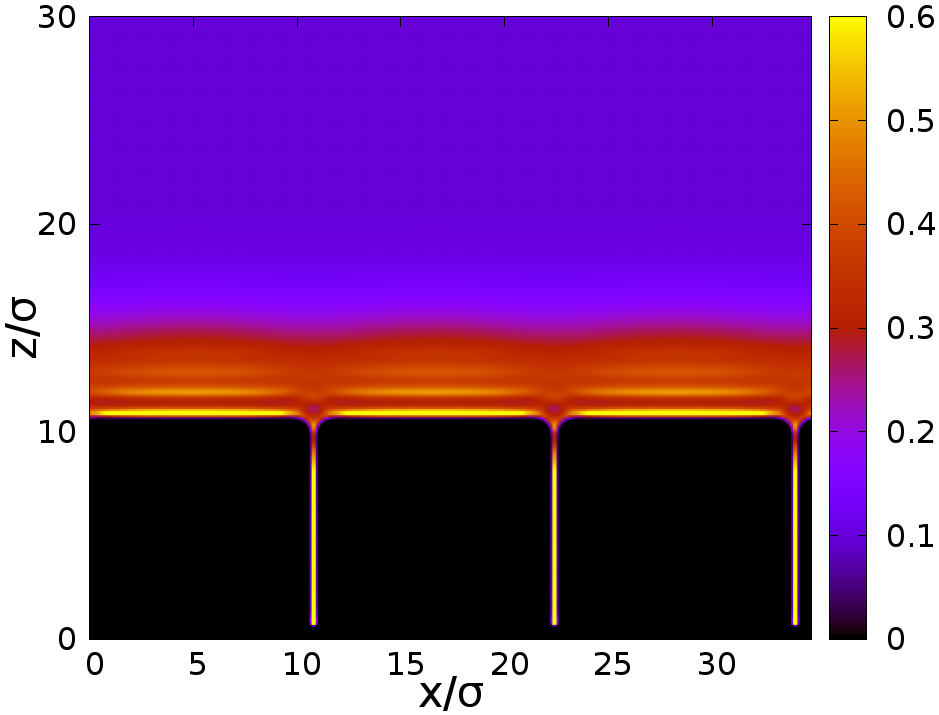}
\includegraphics[width=4.0cm]{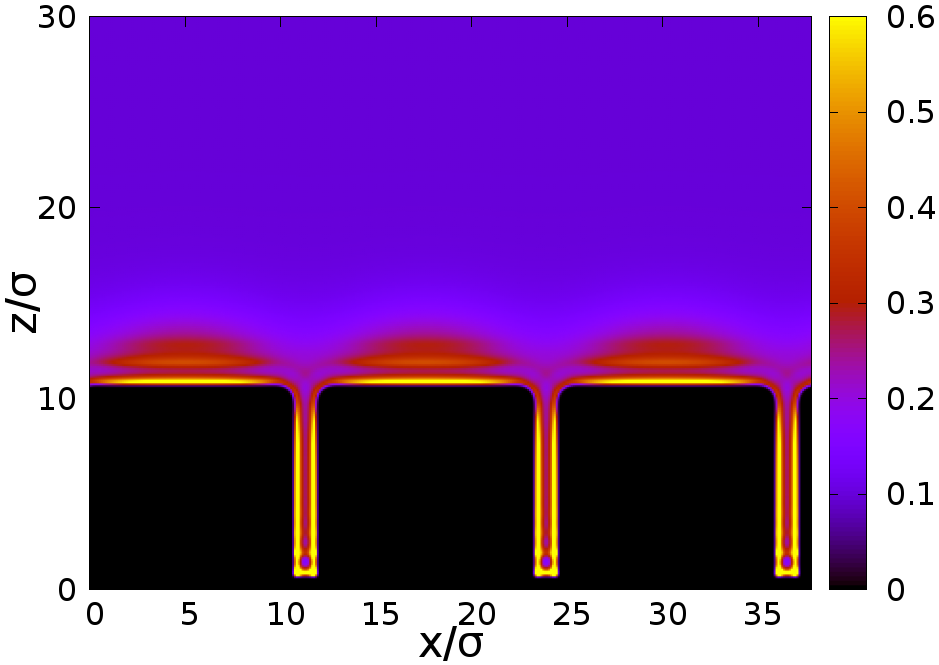} \hspace*{0.1cm} \includegraphics[width=4.0cm]{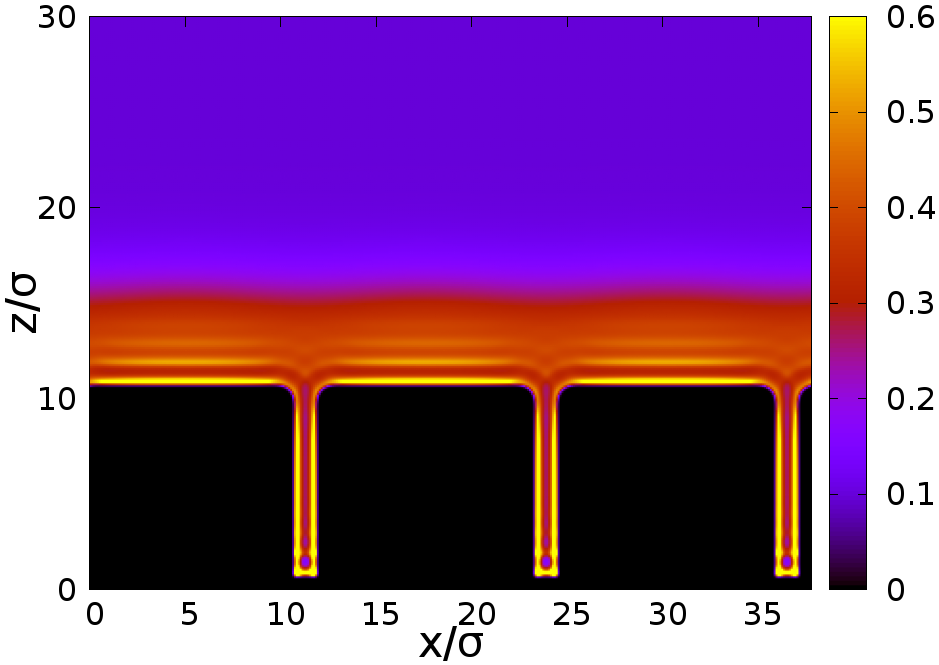}
\includegraphics[width=4.0cm]{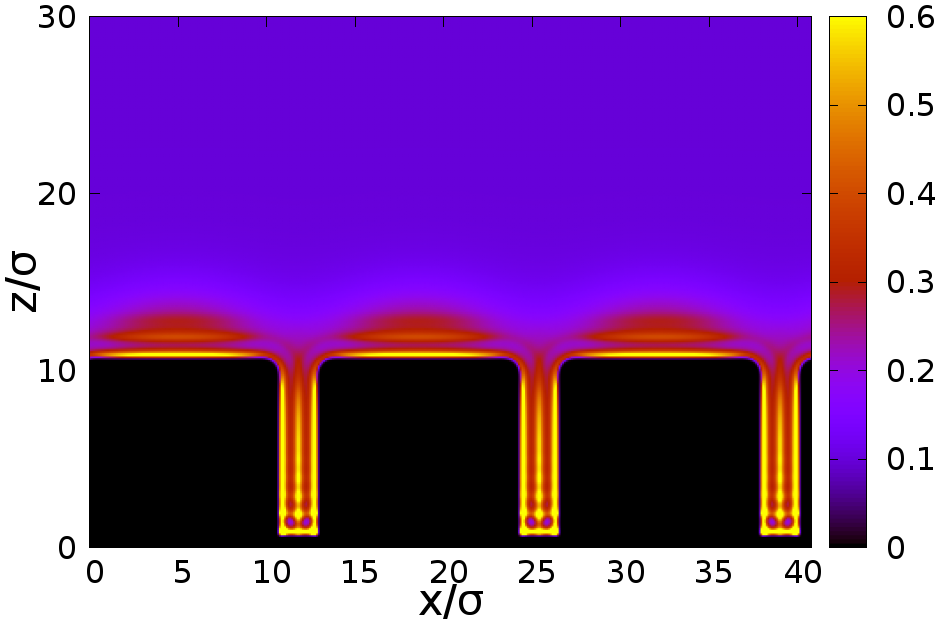} \hspace*{0.1cm} \includegraphics[width=4.0cm]{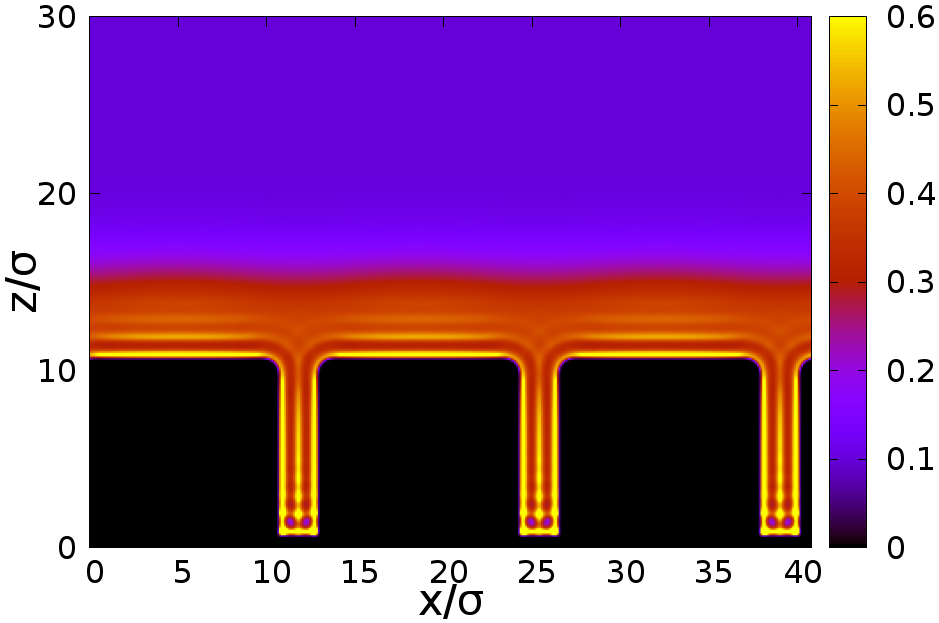}
\includegraphics[width=4.0cm]{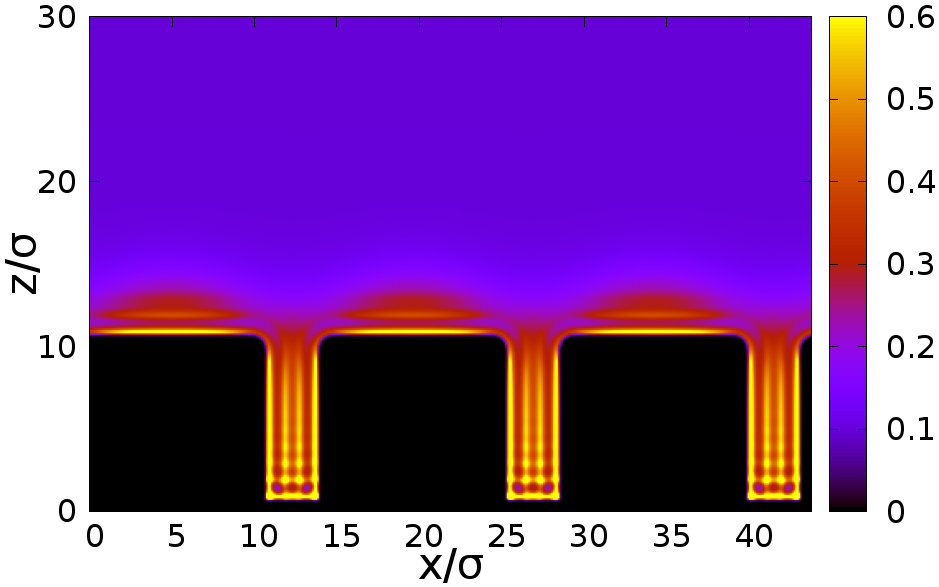} \hspace*{0.1cm} \includegraphics[width=4.0cm]{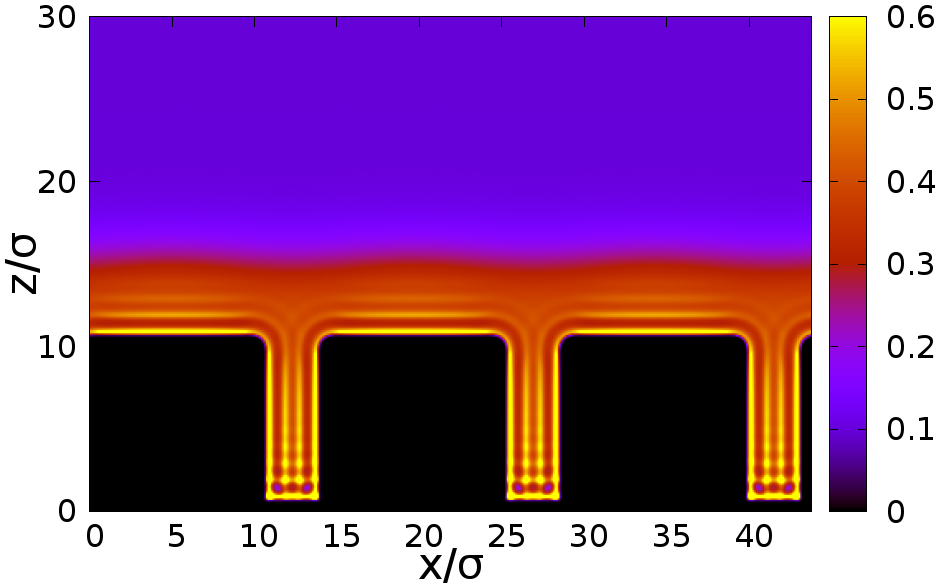}
\caption{Coexisting density profiles of pinned (left panels) and depinned states (right panels) shown over three periods of the wall. The pillar
width is $L_p=10\,\sigma$ in all the cases, while the groove width is (from top to bottom): $L_g=1.6\,\sigma$, $L_g=2.6\,\sigma$, $L_g=3.6\,\sigma$,
and $L_g=4.6\,\sigma$. Note a number of adsorbed layers inside the grooves which gradually increases by one. For $T=0.92\,T_c$.} \label{dep_profs}
\end{figure}

Another interesting feature of the results shown in Fig.~\ref{pd_L2} is that the depinning line exhibits non-monotonic, oscillating behaviour with
increasing amplitudes as $L_g$ decreases. This is a purely microscopic effect due to strong packing effects of the fluid confined in the groove, as
is illustrated by several representative density profiles in Fig.~\ref{dep_profs}. The density profiles correspond to very narrow grooves, the width
of which increases by $1\,\sigma$, which represents the periodicity of the envelope modulating otherwise monotonically increasing character of the
depinning line. The density profiles differ by gradually increasing number of liquid-like layers adsorbed in the grooves, and the strongly
non-monotonic behaviour of the transition line reflects how a certain number of adsorbed layers is commensurate with the given groove width.


In order to interpret al least some of these results we formulate a simple mesoscopic theory describing the depinning transition in terms of the
height of the liquid-gas interface $\ell(x)$ which is a subject of an effective potential arising from the long-range interactions between the fluid
and wall atoms. Before this, it is useful to compare the density profiles of the coexisting phases as is shown in Fig.~\ref{dep_profs}. In general,
the density profiles suggest that in the lower adsorption phase, i.e. before depinning, liquid drops form at pillars and are pinned at the edges.
Avoiding the contributions common for both phases, we can compare the grand potentials per unit length for the lower adsorption (pinned) and the
higher adsorption (depinned) states. The grand potential for the pinned state per unit wall period can be written as
 \bb
  \Omega_{\rm pin}=\sqrt{2A\gamma}\ln L_p+|\delta\mu|\Delta\rho S_{\rm drop} +\tau \label{pin}
 \ee
where $A$ is the Hamaker constant of the corresponding planar wall and $S_{\rm drop}$ is the area of the drop cross-section in the $x$-$z$ plane. The
first contribution to $\Omega_{\rm pin}$ is a logarithmically diverging Casimir-like free energy taking into account surface tension effect and the
substrate potential, and can also be associated with the finite-size scaling of the free-energy at complete wetting \cite{posp17}; the second term is
the free-energy cost due to a presence of an undersaturated liquid and $\tau$ is the line tension associated with the three-phase coexistence at both
pillar edges.

The grand potential corresponding to the depinned state is the global minimum $\Omega_{\rm dep}$ of the function
 \bb
 \tilde{\Omega}_{\rm dep}(\ell)=\frac{AL_p}{(\ell-D)^2}+|\delta\mu|\Delta\rho(\ell-D)(L_p+L_g)\,, \label{dep}
 \ee
 which yields for the equilibrium height $\ell_{\rm eq}$ of the the liquid film:
 \bb
 \ell_{\rm eq}=D+\left[\frac{2AL_p}{|\delta\mu|\Delta\rho(L_p+L_g)}\right]^{\frac{1}{3}}
 \ee
 and upon substituting $\ell_{\rm eq}$ back to (\ref{dep}), we get for the grand potential
 \bb
 \Omega_{\rm dep}=\frac{3}{2}(2AL_p)^{\frac{1}{3}}(|\delta\mu|\Delta\rho(L_p+L_g))^{\frac{2}{3}}\,.  \label{dep2}
 \ee

Here, a number of approximations has been made, namely: 1) In the depinned state, the undulation of the liquid-gas interface is neglected and its
shape is assumed to be flat and of a uniform height $\ell$ measured from the groove bottom; 2) The liquid-gas interface is assumed to be a subject of
the  binding potential $A/h^2$ where $h$ is its normal distance from the wall (this contribution is included explicitly in Eq.~(\ref{dep}) and
implicitly in Eq.~(\ref{pin}) within the first term \cite{posp17}) and the contribution due to the bottom wall is neglected; 3) The shape of the
liquid drops attached at the pillars are deemed not to be appreciably different from that at saturation; therefore, the maximum height of each
droplet is $h_m\approx \sqrt{L_p\sqrt{A/2\gamma}}$  and its local height, when the origin of the coordinate system is put to the middle of a pillar
top,  is $h(x)=h_m\sqrt{1-4(x/L_p)^2}$ \cite{posp17}.  Integrating $h(x)$ along the pillar width yields the area of the drop cross-section $S_{\rm
drop}=\pi/4L_p^{3/2}\left(A/2\gamma\right)^{1/4}$.

Despite its relative simplicity, the mesoscopic model allows to interpret most of the DFT results presented above. First of all, the comparison of
Eqs.~(\ref{pin}) and (\ref{dep}) explains immediately the origin of the depinning transition which resides in the trade-off between the volume and
the surface/interaction terms. More specifically, the greater free-energy cost due to the presence of the metastable liquid in the depinned state is
compensated by the free-energy loss given by narrowing the liquid-gas interface and its greater distance from the wall (the Hamaker constant $A$ is
necessarily positive above the wetting temperature). It means that at low values of the chemical potential (large $|\delta\mu|$) in which case the
volume terms are dominant, the system will be preferentially in the pinned state, while the depinned state becomes more favoured near the bulk
coexistence line, as observed. Furthermore, from the comparison of Eqs.~(\ref{pin}) and (\ref{dep2})  it follows that the depinned state becomes more
stable as $L_p$ is increased, for fixed $\delta\mu$ and $L_g$, due to the positive term in $\Omega_{\rm pin}$ including $S_{\rm drop}\propto
L_p^{3/2}$ which grows fastest with $L_p$, in line with the results shown in Fig.~\ref{pd_L1}. If, on the other hand, we keep $L_p$ fixed and vary
$L_g$ instead, the mesoscopic model predicts that the transition will be shifted towards smaller values of $|\delta\mu|$ with increasing $L_g$, in
view of the increased slope of the free-energy dependence on $|\delta\mu|$ for the depinned state, while the free energy of the pinned state remains
constant (does not depend on $L_g$), which explains the behaviour of the depinning phase boundary shown in Fig.~\ref{pd_L2} for large $L_g$.

Nevertheless, it still remains to clarify the existence of the critical widths $L_{cp}^{-}$ and $L_{cp}^{+}$. As for $L_{cp}^{-}$, i.e. the pillar
width for which the depinning transition occurs at saturation, $\delta\mu=0$, the free energy for the depinned state clearly vanishes, since
$\ell\to\infty$, in contrast to the free energy for the pinned state as given by Eq.~(\ref{pin}). However, for very low values of $L_p$ the pillars
become too narrow to accommodate a liquid  drop in which case the corresponding free energy also vanishes. This suggests that $L_{cp}^{-}$ can be
interpreted as the critical pillar width below which nucleation of liquid drops at the pillars is not possible anymore and for which the microscopic
and macroscopic (complete wetting) adsorption states coexist. This interpretation is supported by the results shown in Fig.~\ref{pd_L1} where all the
depinning transition lines terminate at the value of $L_p\approx4\,\sigma$ independent of $L_g$ and also by the absence of the lower critical point
in the $L_g$--$\delta\mu$ phase diagram shown in Fig.~\ref{pd_L2}.

The mesoscopic model assuming two distinct liquid configurations is unable to predict the critical point $L_{cp}^{+}$ for which the difference
between the pinned and depinned states just disappears. For this, a model containing an order parameter distinguishing between the two states and
vanishing continuously as the critical point $L_{cp}^{+}$ is approached from below is required. However, admitting the existence of the critical
point in the depinning transition, its behaviour observed in Fig.~\ref{pd_L1} can be interpreted rather straightforwardly. Firstly, as the depinning
transition cannot precede the capillary condensation, it follows from Kelvin's equation that the critical point gets closer and closer to the
saturation as the \emph{groove} width $L_g$ is increased in line with the results presented in Fig.~\ref{pd_L1}. Secondly, complementing the free
energy balance by the condition equating the first derivatives of the free energies w.r.t. the chemical potential, we obtain that at the critical
point $L_p\propto {\rm e}^{-\tau/\sqrt{2A\gamma}}$, which explains why, at the given temperature, the critical point occurs at the same value $L_p$
for all the groove widths considered.


\subsection{Unbinding}

\begin{figure}[h]
\centerline{\includegraphics[width=8cm]{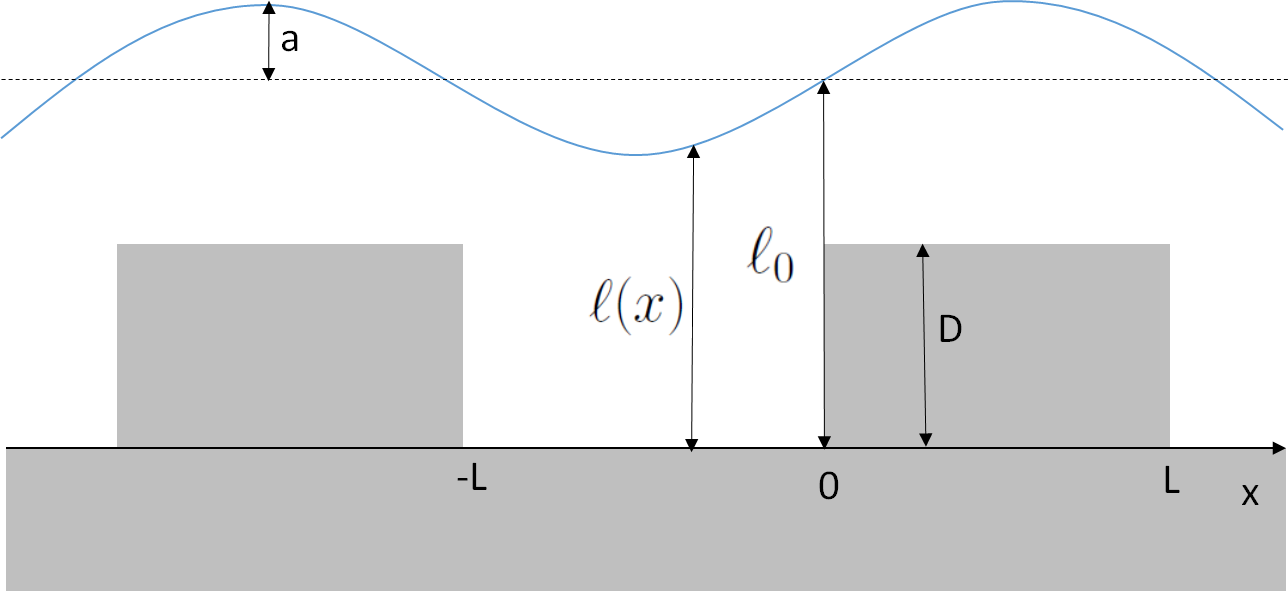}} \caption{Schematic plot showing an anticipated shape of the liquid-gas interface in the unbinding
regime for $L_p=L_g=L$. The local height of the liquid-gas interface $\ell(x)$ is a periodic function with the mean value $\ell_0$ and maximum
$\ell_0+a$. }\label{sketch}
\end{figure}

Finally, we focus on the last adsorption regime which corresponds to unbinding of the liquid-gas interface from the wall. We consider a substrate
with the pillar width $L_p>L_{cp}^-$ and wish to describe the process $\delta\mu\to0^-$  in terms of the  interface shape $\ell(x)$. In what follows
we focus on the ``symmetric'' case, such that $L_p=L_g\equiv L$ illustrated in Fig.~\ref{sketch} and  formulate a simple mesoscopic theory based on
the interface Hamiltonian model which per unit length and one period of the wall can be written as:
 \bb
 H[\ell]=\int_{-L}^{L}\dd x\left\{\frac{\gamma}{2}\left(\frac{\dd \ell(x)}{\dd x}\right)^2
 +W(\ell)\right\} \label{eff_H}
 \ee
where the binding potential $W(\ell)$ is assumed to adopt a simple local form
 \bb
 W(\ell(x))=\left\{\begin{array}{ll}
 |\delta\mu|\Delta\rho\ell(x)+A/\ell(x)^2\,;&x<0\,,\\
|\delta\mu|\Delta\rho(\ell(x)-D)+A/(\ell(x)-D)^2 \,;&x\ge0\,.
\end{array}\right.\label{well}
 \ee
The first term in the binding potential is due to a presence of the metastable liquid and the second term stems from the dispersion interaction
between the liquid-gas interface and the wall. With the the origin of the Cartesian coordinates chosen according to Fig.~\ref{sketch}, the mean
height of the liquid film above the grooves bottom $\ell(x)$ is an odd function with a periodicity $P=2L$ and can thus be reasonably parametrized as
follows
 \bb
 \ell(x)=\ell_0+a\sin\left(\frac{\pi x}{L}\right) \label{param}
 \ee
 where we expect $a\ll\ell_0$ for small $\delta\mu$.

 Substituting (\ref{well}) and (\ref{param}) into (\ref{eff_H}) leads to
  \begin{eqnarray}
  H(a,\ell_0)&=&\frac{\gamma}{2}\int_{-L}^L\left(\frac{\dd \ell(x)}{\dd x}\right)^2\dd x+|\delta\mu|\Delta\rho L(2\ell_0-D)\nonumber\\
  &&+\int_{-L}^0\frac{A}{\ell^2}\dd x+\int_0^{L}\frac{A}{(\ell-D)^2}\dd x \nonumber\\ 
  &=&\frac{\gamma}{2}\frac{1}{L}\int_{-1}^1\left(\frac{\dd \ell(\tilde{x})}{\dd \tilde{x}}\right)^2\dd \tilde{x}
  +|\delta\mu|\Delta\rho L(2\ell_0-D)\nonumber\\
  &+&AL\int_{-1}^0\frac{1}{\ell^2(\tilde{x})}\dd \tilde{x}+AL\int_0^{1}\frac{1}{(\ell(\tilde{x})-D)^2}\dd \tilde{x}\,.\nonumber\\
  &=&\frac{\gamma a^2\pi^2}{2L}+|\delta\mu|\Delta\rho L(2\ell_0-D)\label{htild}\\
  &+&AL\int_{-1}^0\frac{1}{\ell^2(\tilde{x})}\dd \tilde{x}+AL\int_0^{1}\frac{1}{(\ell(\tilde{x})-D)^2}\dd \tilde{x}\,.\nonumber
  \end{eqnarray}
  The integrals can be carried out analytically
  \begin{eqnarray}
\hspace*{-0.7cm} \int_{-1}^0\frac{\dd
 \tilde{x}}{\ell^2(\tilde{x})}&=&\frac{\pi\ell_0^2+2\ell_0^2\arctan\left(\frac{a}{\sqrt{\ell_0^2-a^2}}\right)
 +2a{\sqrt{\ell_0^2-a^2}}}{\pi\ell_0(\ell_0^2-a^2)^{\frac{3}{2}}}\nonumber\\
 &=&\frac{1}{\ell_0^2}+\frac{4}{\pi}\frac{a}{\ell_0^3}+{\cal{O}}(a^2/\ell_0^2)\,,
 \end{eqnarray}
 considering small $\delta\mu$.
 Similarly,
 \begin{eqnarray}
\int_{0}^1\frac{\dd \tilde{x}}{(\ell(\tilde{x})-D)^2}&=&\frac{1}{(D-\ell_0)^2}-\frac{4}{\pi}\frac{a}{(\ell_0-D)^3}\nonumber\\
&&+{\cal{O}}(a^2/(\ell_0-D)^2)\,.
 \end{eqnarray}
 To first order in $a/\ell_0$ and $a/(D-\ell_0)$, $H(a,\ell_0)$ becomes
  \begin{eqnarray}
  H(a,\ell_0)&\approx&\frac{\gamma a^2\pi^2}{2L}+|\delta\mu|\Delta\rho L(2\ell_0-D)\label{H_bent}\\
  &+&AL\left[\frac{1}{\ell_0^2}+\frac{4}{\pi}\frac{a}{\ell_0^3}+\frac{1}{(\ell_0-D)^2}-\frac{4}{\pi}\frac{a}{(\ell_0-D)^3}\right]\,.\nonumber
\end{eqnarray}
 At a given $\delta\mu$, the equilibrium state corresponds to the stationary point:
 \begin{eqnarray}
 \frac{\partial H}{\partial a}=\frac{\partial H}{\partial \ell_0}=0 \,,  
 \end{eqnarray}
 implying
  \bb
|\delta\mu|=\frac{A}{\Delta\rho}\left[\frac{1}{\ell_0^3}+\frac{1}{(\ell_0-D)^3}+\frac{6a}{\pi\ell_0^4}
 -\frac{6}{\pi}\frac{a}{(\ell_0-D)^4}\right]\label{dHdl}
\ee
 and
  \bb
  a\approx \frac{12AL^2}{\pi^3\gamma}\frac{D}{\ell_0^4}\,.\label{dHda}
  \ee

\begin{figure}[h]
\centerline{\includegraphics[width=8cm]{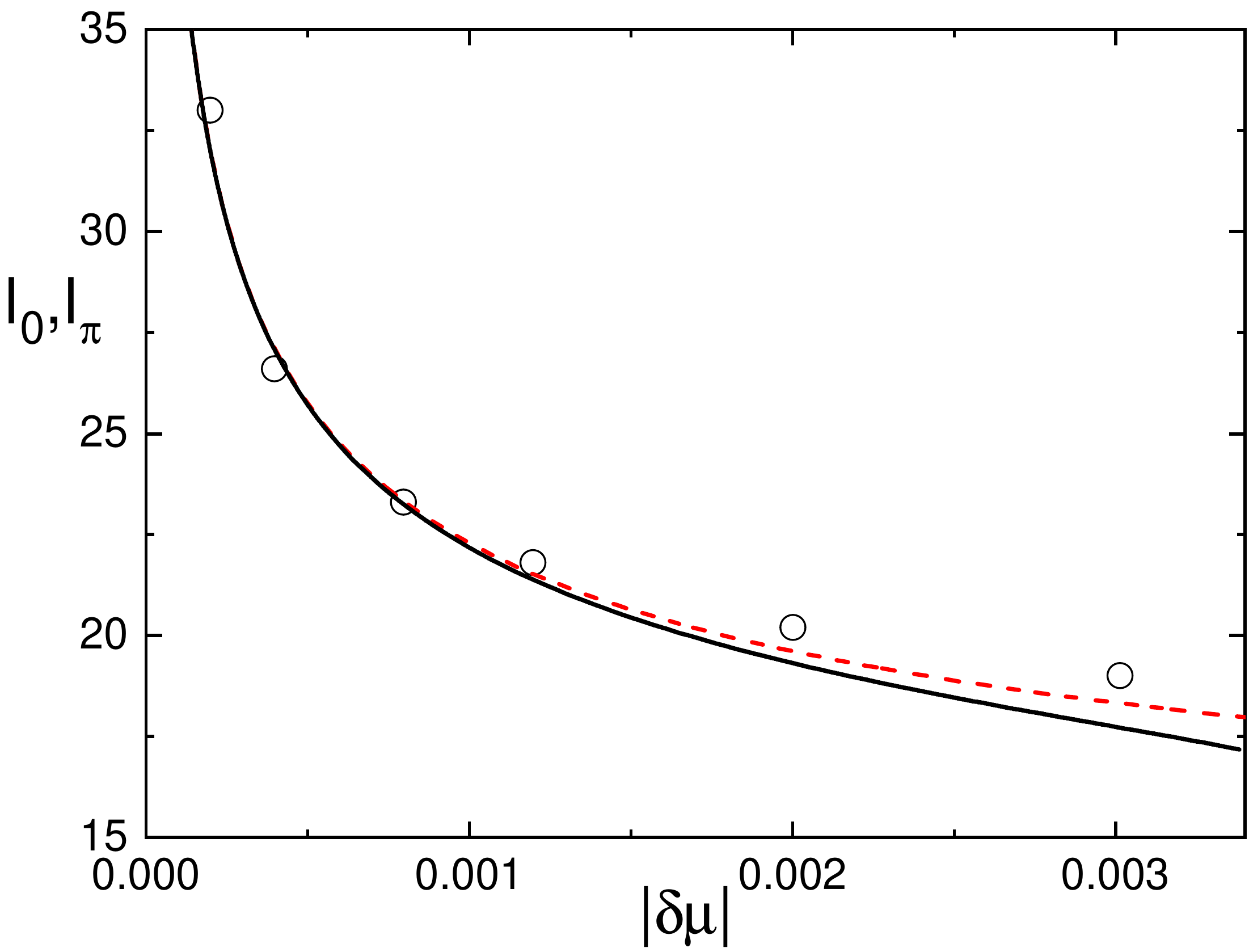}} \caption{Mean height of the liquid-vapour interface (in units of $\sigma$) as a function of
$|\delta\mu|=\mu_{\rm sat}-\mu$ (in units of $\varepsilon$) measuring the departure of the given thermodynamic state from the bulk two-phase
coexistence. The solid line represents the prediction based on the interfacial Hamiltonian theory as given by equations (\ref{dHda}) and
(\ref{dHdl}), while the symbols represent DFT results. Also shown (red dashed line) is the interface height corresponding to configurations with a
flat interface. For the temperature $T=0.92\,T_c$. }\label{p_ell}
\end{figure}

\begin{figure}[h]
\centerline{\includegraphics[width=9cm]{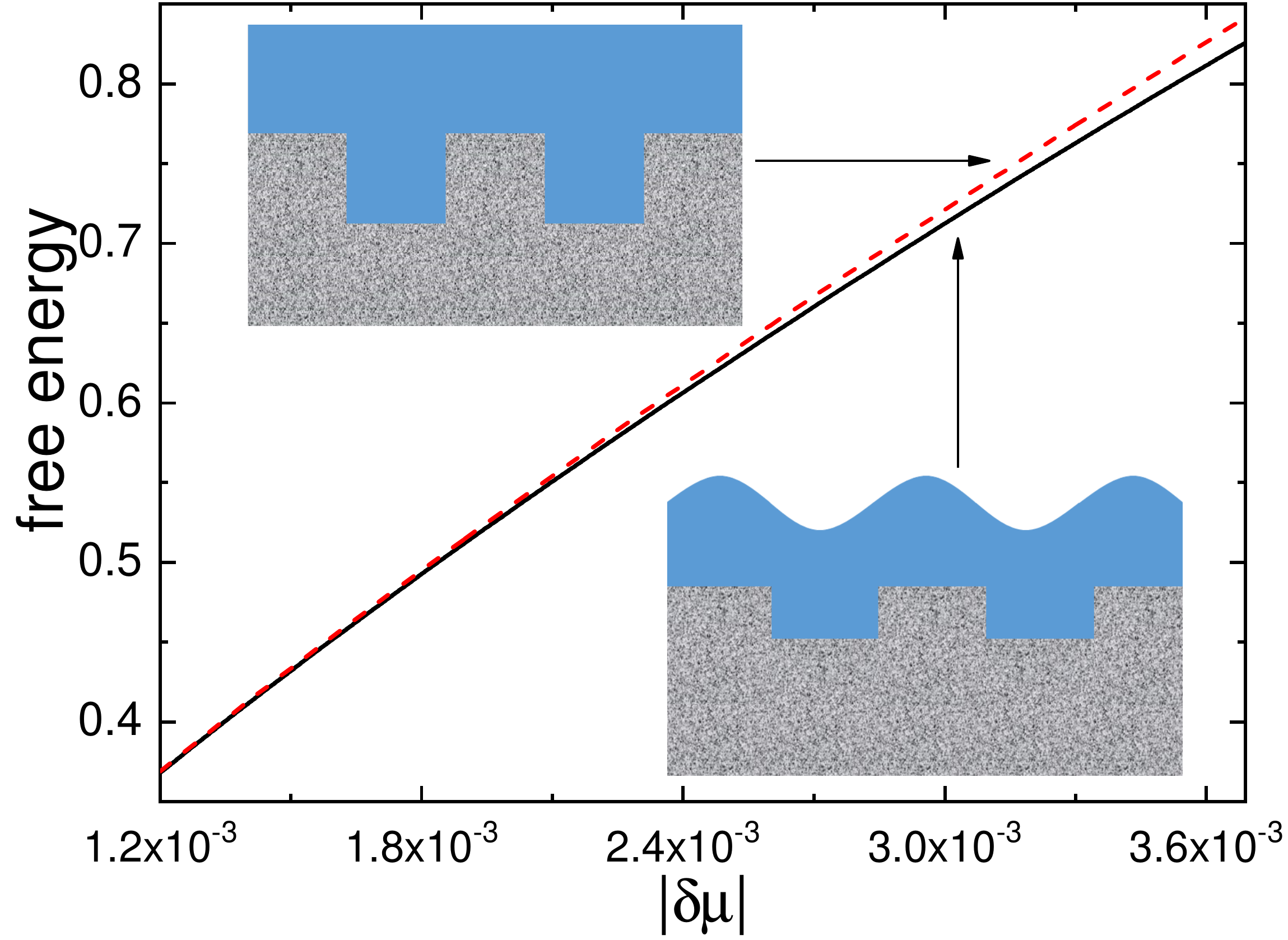}} \caption{A comparison of the free energy per unit length (in units of $\varepsilon/\sigma$) for the
undulated (sinusoidal) and flat configurations as determined from Eqs.~(\ref{F_bent}) and (\ref{F_unbent}), respectively, as a function of the
chemical potential difference from saturation $|\delta\mu|$ (in units of $\varepsilon$).}\label{p_F}
\end{figure}

\begin{figure}[h]
\includegraphics[width=9cm]{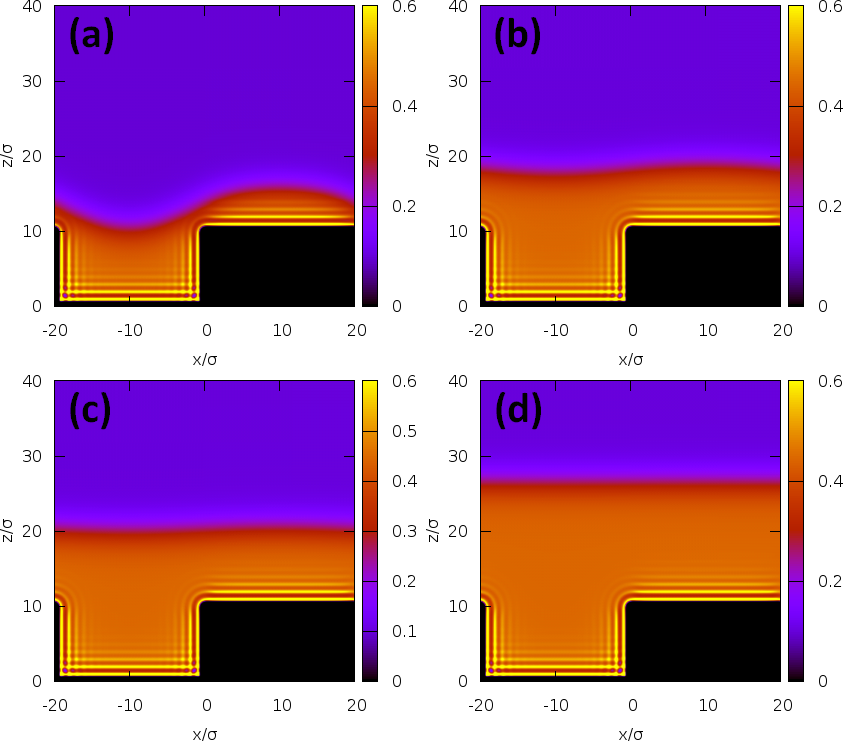}
\caption{Two-dimensional DFT density profiles shown over a single period of the wall with the groove depth of $D=10\,\sigma$ and the pillar/groove
width of $L=20\,\sigma$.  The density profiles correspond to the chemical potential departure from saturation (in units of $\varepsilon$):
 (a)  $2\cdot10^{-2}$, (b) $4\cdot10^{-3}$, (c)  $2\cdot10^{-3}$, and  $4\cdot10^{-4}$. For the temperature $T=0.92\,T_c$. } \label{pT13}
\end{figure}

It may be interesting to compare the mean height $\ell_0$ of the equilibrium undulated interface with that of a flat interface $\ell_\pi$ where we
simply set $a=0$ and which is obtained by minimizing the Hamiltonian function
 \bb
 H_\pi(\ell)=|\delta\mu|\Delta\rho L(2\ell-D)+\frac{AL}{\ell^2}+\frac{AL}{(\ell-D)^2}\,,
 \ee
which yields
  \bb
  |\delta\mu|=\frac{A}{\Delta\rho}\left[\frac{1}{\ell_\pi^3}+\frac{1}{(\ell_\pi-D)^3}\right]\,. \label{lpi}
  \ee

\begin{figure}[h]
\centerline{\includegraphics[width=8cm]{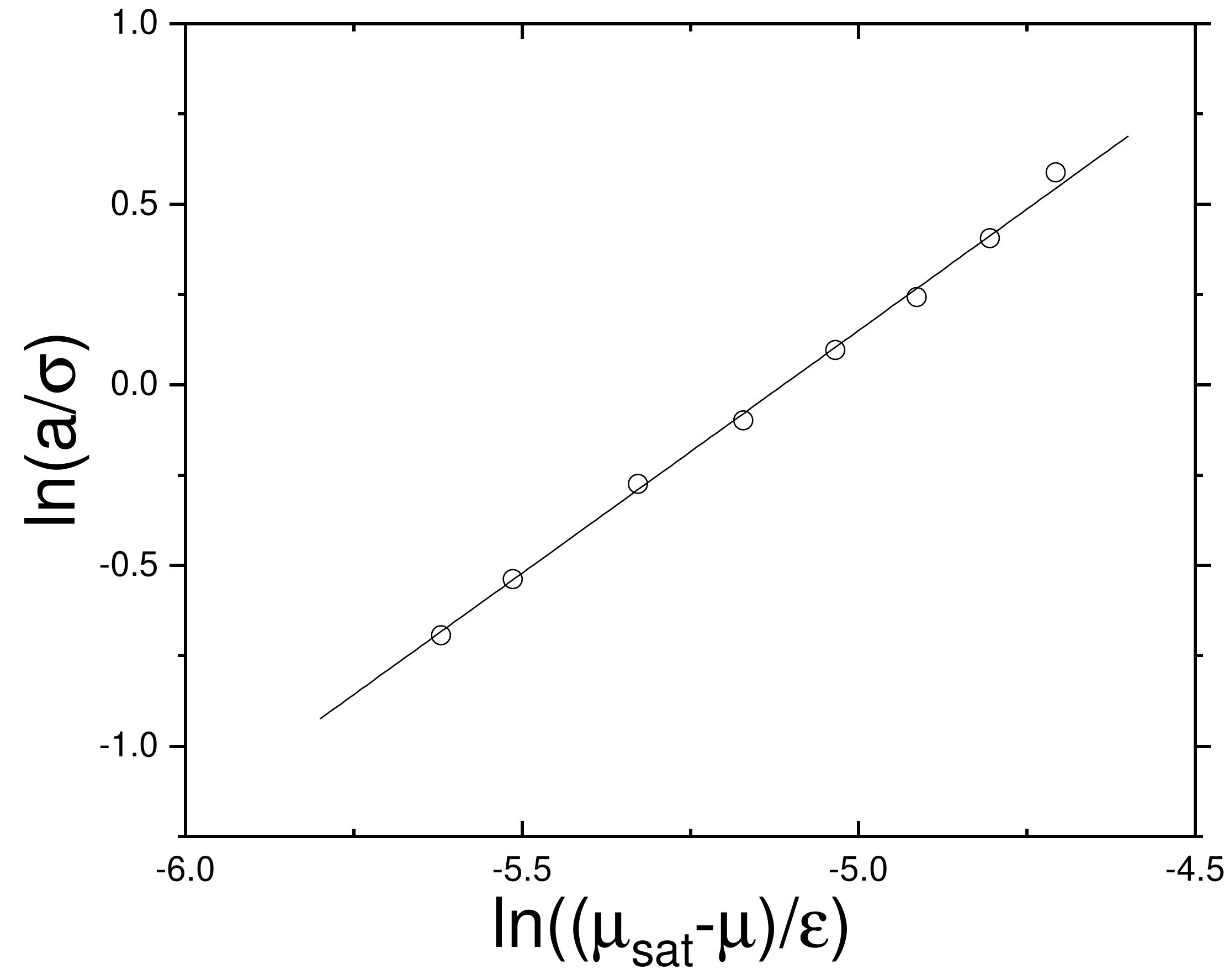}}
 \caption{A log-log plot of the dependence of the interface amplitude $a$ on the chemical potential
departure from saturation as obtained from DFT for the substrate with $D=10\,\sigma$ and $L=20\,\sigma$ at temperature $T=0.92\,T_c$. The slope of
the fitting line is approximately $1.35$.} \label{fig14}
\end{figure}

In Fig.~\ref{p_ell} we compare the mean height of the undulated interface, $\ell_0$, as obtained by solving  Eqs.~(\ref{dHdl}) and (\ref{dHda}), with
the height of the flat interface with $\ell_\pi$, obtained from Eq.~(\ref{lpi}). Also shown are the DFT results of $\ell(0)$ determined from the
density profiles using  the mid-density rule, $\rho(0,\ell(0))=(\rho_v+\rho_l)/2$, for the substrate with $D=10\,\sigma$ and $L=20\,\sigma$.  The
comparison reveals that all the three sets of results are very close to each other in the displayed range of the chemical potential and essentially
collapse to a single curve as $|\delta\mu|<10^{-3}\varepsilon$. We can thus conclude that the interface Hamiltonian model in a combination with the
parametrization (\ref{param}) provides an accurate prediction for the mean height of the unbinding interface $\ell_0$ which, moreover, is shown to
get increasingly close to that of the flat interface $\ell_\pi$, as the saturation is approached, hence $\ell_0\sim|\delta\mu|^{-1/3}$, as
$|\delta\mu|\to0$. Furthermore, from this and Eq.~(\ref{dHda}) it follows for the interface roughness $a\sim\delta\mu^{4/3}$, as $|\delta\mu|\to0$.

In order to obtain still more detailed insight into the process of the interface flattening, we compare the free energies pertinent to the undulated
and flat interfaces. They are, respectively, given by
 \bb
 F=2L\gamma+{\rm min}\{H(\ell_0,a)\} \label{F_bent}
 \ee
 and
 \bb
 F_\pi=2L\gamma+{\rm min}\{H_\pi(\ell_\pi)\} \label{F_unbent}
 \ee
where the contribution $2L\gamma$ due to the free (flat) liquid-gas interface has been included. Assuming that $\ell_0\approx\ell_\pi$, as justified
for small $\delta\mu$ according to the previous results, the free-energy difference is
 \begin{eqnarray}
 \Delta F&=&F-F_\pi\approx\frac{\pi^2\gamma a^2}{2L}+\frac{4ALa}{\pi}\left[\frac{1}{\ell_0^3}-\frac{1}{(\ell_0-D)^3}\right]\nonumber\\
         &\approx&-\frac{\pi^2\gamma a^2}{2L}\,.
 \end{eqnarray}

The comparison of the free energies shown in Fig.~\ref{p_F} reveals that the equilibrium interface flattens in a continuous way, such that the free
energy difference between the undulated and flat states decays  as $\Delta F\sim\ell_0^{-8}\propto|\delta\mu|^{8/3}$, as $\delta\mu\to0$. In
Fig.~\ref{pT13}, density profiles as obtained from DFT for the substrate with $D=10\,\sigma$ and $L=20\,\sigma$ are shown  to illustrate the
interface flattening as the saturation is approached. From the density profiles, the amplitude $a$ of the interface can be determined using the same
mid-density rule as described above. In Fig.~\ref{fig14} the log-log plot of the dependence of $a$ on $\delta\mu$ is displayed which obeys a linear
behaviour in line with the expected power-law dependence. Moreover, the line with the best fit to the data has a slope which is very close to the
predicted value of $4/3$.



\section{Summary and concluding remarks}

In this work we studied complete wetting ($T>T_w$) of periodically structured substrates which interact with the fluid via long range (dispersion)
forces. The model substrates are formed by scoring rectangular, macroscopically long grooves of depth $D$ and width $L_g$ into a planar wall, such
that the grooves are separated by pillars of width $L_p$.  The entire adsorption  process can be divided into three parts: (i) the filling regime
within which the grooves become filled with liquid; (ii) the depinning regime which corresponds to (continuous or discontinuous) merging of the
liquid columns filling the grooves and (iii) the unbinding regime when the (single) liquid-gas interface moves away from the substrate and its
undulated shape gradually flattens. All the three regimes have been examined separately in some detail and the main conclusions can be summarized as
follows:

\begin{itemize}

\item \emph{Filling regime}: In the initial part of adsorption (well below $\mu_{\rm sat}$), the liquid phase nucleates at the bottom of the grooves,
while the upper part of the wall (pillars) are covered by only a microscopically thin layer of liquid. In contrast to low temperatures $T<T_w$, in
which case the grooves become filled with liquid via a first-order transition,  the filling process is continuous for temperatures $T>T_w$ and can be
associated with a gradual rise of menisci separating liquid and gas phases in each groove, as the chemical potential is increased. For sufficiently
deep grooves one can identify a critical exponent $\beta_g$ characterizing the growth of each meniscus as the chemical potential approaches the value
$\mu_{cc}(L_g)$, pertinent to capillary condensation in an infinite slit of the same width and at the same temperature. However, compared to
thoroughly studied systems involving only a single groove, the filling process for the current model with a periodic array of grooves deviates in two
ways: Firstly, the power-law dependence for the height of the meniscus is now associated with the difference of the chemical potential from the value
$\mu_{cc}^{L_p}(L_g)$ which corresponds to capillary condensation inside a slit formed of a pair of walls of width $L_p$ and which is higher than
$\mu_{cc}(L_g)$, in view of the weaker wall potential. Secondly, for the similar reasons, the critical exponent is now $\beta_g=1/3$ rather than
$1/4$ as valid the for the single groove, since the grooves are now effectively chemically heterogenous (due to the weaker effective strength of the
side walls potential) -- this implies that the fine compensation of the leading order terms in the binding potential is not present anymore (as for
the single groove) which results in the shift of the denominator in $\beta_g$ by one.

If the grooves are only microscopically deep, the filling process is no more critical and cannot thus be characterized by a critical exponent anymore
but there is a new aspect instead. Now, the effective repulsive  potentials induced by the groove bottom and groove top which both repel the meniscus
compete with each other and may give rise to a localization-delocalization transition. For the walls which exhibit first-order wetting transition, as
considered here, the localization-delocalization transition is also first-order and induces a jump in the height of the meniscus at capillary
liquid-gas coexistence $\mu_{cc}(L_g)$. The transition terminates at temperature $T_s$ above which the corresponding binding potential possesses only
a single minimum.

\item \emph{Depinning regime}: After the grooves get filled with liquid, i.e. the menisci reach the grooves top, a single liquid-gas interface
eventually forms within the regime which is referred to as depinning. As already pointed out recently, this process is either first-order or
continuous depending on the wall parameters. Here, we have shown that the nature of the depinning depends solely on the pillar width $L_p$ which
possesses two threshold values $L_{cp}^-$ and $L_{cp}^+$, such that: for $L>L_{cp}^+$ the process is continuous, for $L_{cp}^-<L<L_{cp}^+$ the
depinning is first-order transition and for $L<L_{cp}^-$ the depinning does not occur and the wall remains non-wet at saturation. The values
$L_{cp}^-$ and $L_{cp}^+$ are microscopically small and are independent of the grooves width $L_g$ which only determines the location of the
depinning transition (if present). The dependence of the location of the transition on the groove width is non-monotonic and exhibits strongly
oscillating character for small values of $L_g$ with a periodicity of one molecular diameter due to packing effects that induce well distinguishable
liquid layers inside the grooves. Although microscopic in nature, most of these phenomena observed using the microscopic density functional theory
can be explained by a simple mesoscopic theory. In particular,  $L_{cp}^-$ has been interpreted as a critical pillar width below which a liquid drop
cannot be accommodated, while $L_{cp}^+$ has been shown to be tied with the line tension.

\item \emph{Unbinding regime}: Finally, for any value of the pillar width $L>L_{cp}^-$ a single liquid-gas interface forms and unbinds from the
wall as $\mu$ approaches $\mu_{\rm sat}$. In contrast to complete wetting on a planar wall, the liquid gas interface in now periodically undulated
(even on a mean-field level) but its mean height $\ell_0$ grows in a very similar fashion to that of a flat interface and eventually diverges in the
limit of $\delta\mu\to0$ according to the same power-law, i.e. with the critical exponent $\beta_{co}=1/3$ for systems involving dispersion forces.
The growth of the interface is simultaneously accompanied by a gradual flattening of the interface (or unbending), such that the undulation amplitude
decays as $a\sim \ell_0^{-4}$, or, alternatively, as $a\sim |\delta\mu|^{4/3}$ for ``symmetrically'' structured substrates, such that $L_g=L_p$.

In conclusion, we have seen that despite its relatively simple structure, the model of a grooved substrate predicts an interesting interplay of
various surface phase transitions giving rise to a very complex phase behaviour of adsorbed fluids. The presence and nature of the phase transitions
depend sensitively on the wall parameters, such that the entire adsorption isotherm may reflect a whole sequence of phenomena: within the filling
regime a first-order localization-delocalization can occur for shallow grooves or a continuous filling transition for deep grooves; this is followed
by a depinning transition which can be rounded, critical or first-order depending on the width of the pillars separating the grooves; finally, the
adsorption isotherm diverges continuously due to complete wetting of the top surface which, moreover, can still be preceded be a prewetting jump (not
shown here). Some of the conclusions deserve further comments. For instance, we have claimed that the critical exponent for the filling in deep
grooves is now $\beta_g=1/3$, rather than $\beta_g=1/4$ pertinent to a single groove. This conclusion was supported by DFT results considering a
rather extreme case of very thin pillars of width $L_p=2\,\sigma$, while the width of the grooves was $L_g=10\,\sigma$. However, for substrates with
very thick pillars, such that $L_p\gg L_g$, it can be anticipated that filling does not appreciable differ from that present in a single groove and
therefore the critical exponent $\beta_g$ may actually interpolate between $1/4$ and $1/3$ in this case. In fact, these arguments are only of a
mean-field character and $\beta_g$ ultimately crossovers to the true value $\beta_g=1/3$ for $\mu$ very close to $\mu_{cc}^{L_p}(L_g)$ in any case,
due to fluctuation effects \cite{our_groove}. The capillary fluctuations have even more significant effect regarding a possible finite jump in the
height of the meniscus, since this phenomenon occurring individually in each groove is a pseudo-1D transition and is thus expected to be rounded in a
real experiment. On the other hand, the depinning transition is a phenomenon where the grooves (and pillars) are collectively involved and thus will
not be destroyed by fluctuation effects.

Clearly, the work can be extended by numerous modifications of the substrate model. One can consider different geometries of the pits which will
presumably have a strong impact on the nature of the phase transitions; for example, one does not expect a presence of the depinning transition for
smooth surfaces that do not involve sharp edges. One can also attempt to obtain a more realistic approximation for rough solid surfaces by
considering grooves of different depths, widths or separations; condensation in grooves formed by two differently high side walls is actually of some
interest in its own right. Finally, one may further decorate the surface of the wall and investigate the effect of such smaller length-scale defects.
We intend to study such problems, some of which require a 3D DFT analysis, in future work.

\end{itemize}

\begin{acknowledgments}
\noindent This work was financially supported by the Czech Science Foundation, Project No. GA 20-14547S and the European Union's Horizon 2020
research and innovation program (Project VIMMP: Virtual Materials Marketplace, No. 760907).
\end{acknowledgments}

\end{document}